\documentclass{aa}  

\usepackage{txfonts}
\usepackage{graphicx,lipsum,overpic}
\usepackage{subfigure}
\usepackage{amssymb,amsmath}

\newcommand{\ignore}[1]{}

\def\aap {{A\&A}}

\def\apjl {{ApJL}}

\def\apjs {{ApJS}}

\begin{document}

   \title{Connecting planet formation and astrochemistry }

  \subtitle{Refractory carbon depletion leading to super-stellar C/O \\ in giant planetary atmospheres}

   \author{A.J. Cridland\inst{1}\thanks{cridland@strw.leidenuniv.nl}, Christian Eistrup\inst{1}, \& Ewine F. van Dishoeck\inst{1}
          }

   \institute{
   $^1$Leiden Observatory, Leiden University, 2300 RA Leiden, the Netherlands
             }

   \date{Received \today}


  \abstract
  {
Combining a time-dependent astrochemical model with a model of planet formation and migration, we compute the carbon-to-oxygen ratio (C/O) of a range of planetary embryos starting their formation in the inner solar system (1-3 AU). Most of the embryos result in Hot Jupiters (M $\geq$ M$_{J}$, orbital radius $<$ 0.1 AU) while the others result in super-Earths at wider orbital radii. The volatile and ice abundance of relevant carbon and oxygen bearing molecular species are determined through a complex chemical kinetic code which includes both gas and grain surface chemistry. This is combined with a model for the abundance of the refractory dust grains to compute the total carbon and oxygen abundance in the protoplanetary disk available for incorporation into a planetary atmosphere. We include the effects of the refractory carbon depletion that has been observed in our solar system, and posit two models that would put this missing carbon back into the gas phase. This excess gaseous carbon then becomes important in determining the final planetary C/O because the gas disk now becomes more carbon rich relative to oxygen (high gaseous C/O). One model, where the carbon excess is maintained throughout the lifetime of the disk results in Hot Jupiters that have super-stellar C/O. The other model deposits the excess carbon early in the disk life and allows it to advect with the bulk gas. In this model the excess carbon disappears into the host star within 0.8 Myr, returning the gas disk to its original (sub-stellar) C/O, so the Hot Jupiters all exclusively have sub-stellar C/O. This shows that while the solids will tend to be oxygen rich, Hot Jupiters can have super-stellar C/O if a carbon excess can be maintained by some chemical processing of the dust grains. The atmospheric C/O of the super-Earths at larger radii are determined by the chemical interactions between the gas and ice phases of volatile species rather than the refractory carbon model. Whether the carbon and oxygen content of the atmosphere was accreted primarily by gas or solid accretion is heavily dependent on the mass of the atmosphere and where in the disk the growing planet accreted.
}


    \maketitle
%
\section{ Introduction }\label{sec:intro}

The carbon-to-oxygen elemental ratio (C/O) in a planetary atmosphere will be dictated by the chemical properties of its natal protoplanetary disk, the details of which are complicated by chemical evolution \citep{Eistrup2016,Eistrup2017,Crid17}, and planet formation history \citep{Madu2014,Madhu2017,Thiabaud2015,Mordasini16,Crid16a}. However, C/O is our best tracer of planet formation because the elemental budget of carbon and oxygen have been well studied in the diffuse ISM \citep{Whittet2013,Boogert2015} and in protoplanetary disks \citep{Pon14}. Moreover, its variation through the disk (see \citealt{Oberg11} and \citealt{Eistrup2017}) can give us clues to the location and timing of a planet's atmospheric accretion \citep{Crid16a,Madhu2017}. 

Through gas accretion alone, \cite{Crid17} showed that planets tend to have atmospheric C/O that exactly matches the initial C/O of the gas disk inward of the water ice line. This result depends heavily on the physical details that dictate the rate of planetary migration. In particular, \cite{Crid17} invoke `planet-traps' to solve the `Type-I migration problem' - where planets with mass $\le$ few M$_\oplus$ migrate into the host star faster than they grow. These traps (also called convergence zones by \cite{Mordasini2011} and \cite{Horn2012}) are discontinuities in the physical properties of the disk (ie. dust opacity, turbulence strength, heating mechanism) that produce a converging zero-point in the net torque \citep{Masset06,HP13,Cole14,Cole16,Crid16a}. 

In planet trap models the migration of the growing planet is restricted to the radial evolution of the trap, and in the case of \cite{Crid17} each of the traps converge at the water ice line late in the disk life. Coincidentally each of the growing planets were brought to the water ice line before they began accreting their gas. Water freezes out at the highest temperature of any volatile, and hence near the water ice line all volatiles are in their gas phase, and the resulting C/O of the planet is unchanged relative to the C/O of the volatiles (gas and ice) that are inherited from the pre-stellar core. So the combination of converging planet traps near the water ice line and the chemistry of the gas near the ice line is what led to the ubiquitous result reported in \cite{Crid17}.

While important, gas accretion may not be the sole contributor to the metallicity (and C/O) of an exoplanetary atmosphere. Indeed both \cite{Thiabaud2015} and \cite{Mordasini16} demonstrate that when icy planetesimals accrete directly into the forming atmosphere they can evaporate (as computed by \cite{Mordasini2006}, \cite{Mordasini15}) and deliver their ices and refractories (ie. solid carbon and silicates) directly into the planetary atmospheres. Assuming efficient mixing, this mechanism would enrich the atmosphere in carbon, oxygen, and the other elements that make up the planetesimal. This could also be a plausible explanation for Jupiter's apparent carbon enrichment relative to the Sun \citep{Atreya2016}.

The chemistry of the disk gas and ice is commonly treated simply in planet formation models. The radial distribution of the carbon and oxygen content of the gas and ice have previously been prescribed by a constant step-function (as in \citealt{Oberg11}). Each outward step is located at the ice line of the most common volatiles and are caused by volatile freeze out, transferring carbon and/or oxygen from the gas to the ice phase. In this way, the position of a planet's gas accretion relative to the ice lines in the disk midplane dictates its C/O.

An extra level of complexity can be added by considering the radial evolution of the gas and dust. In principle any radial feature in the chemistry of the gas will advect inward with the bulk gas \citep{Thiabaud2015}. However the dust also radially drifts inwards faster than the gas \citep{W77}, which transports volatile species across their ice line \citep{Booth2017,Bosman2017b} leading either to an enhancement or maintenance of the gaseous C/O against inward advection. Here we will focus mostly on the effect of chemical evolution on the resulting C/O of planetary atmospheres, and will ignore the effect of radial evolution on the abundance of the carbon and oxygen that is incorporated in volatile species (either gas or ice).

For the carbon and oxygen coming from refractory species (solid carbon and silicates, incorporated into planetesimals), we generally assume that they stay chemically inert. The exception being our model of refractory carbon depletion that is based on observations of the carbon content in rocky bodies in our own solar system \citep{Allegre2001,Lee2010,Berg15}. Both asteroids in the main belt and our Earth show signs of carbon depletion by between 2 and 3 orders of magnitude respectively, relative to the interstellar medium. As such \cite{Mordasini16} included a simple model for carbon depleted refractories in their planet formation model which contributed to their findings that planetary atmospheres tend to have sub-stellar C/O. In this work we similarly include refractory carbon depletion in the solids, however we additionally put this missing carbon back into the gas phase - allowing it to be accreted into a planetary atmosphere.

It is a common practice to compare observed chemical abundances in (exo)planetary systems to the photosphere of their host star, as it is assumed that the stellar photosphere acts as a time capsule for the chemical composition of the gas and refractory material that it accretes through its protoplanetary disk. Here we assume that the C/O of the stellar photosphere represents all of the carbon and oxygen that was accreted by the star over the course of its formation. Such an assumption is most relevant for stars with smaller convective cells at the top of their atmosphere so that the gas is not mixed with the inner core. Comparing the observed planetary and host star C/O is thus an important observational tool for understanding the process of planet formation. A current observational survey of host star and planet C/O done by \cite{Brewer16} have suggested that planetary C/O tend to be super-stellar.

\ignore{ These sub-stellar C/O results discussed above are currently at odds with observational work that suggest that planetary atmospheres have either stellar or super-stellar C/O \citep{Brewer16}. }

In this work we combine planet formation models with the chemical results from simulations using a chemical kinetics code (the BADASS-code, see \cite{McE03}, \cite{Walsh15}, and references therein), which combines both gas and grain-surface chemistry using an extensive network of reactions, and to this we add two models for returning refractory carbon to the gas. In this way we study the important contributors to the delivery of carbon and oxygen as well as how the temporal evolution of the disk C/O is imprinted onto the chemical structure of an exoplanetary atmosphere. In \S \ref{sec:CtoO} we outline how the carbon and oxygen is distributed in the disk, including a carbon excess coming from the refractory dust. In \S \ref{sec:chemical} we outline the chemical and disk model, while in \S \ref{sec:plntform} the planet formation model is outlined. We outline and discuss our results in \S \ref{sec:results}, and conclude in \S \ref{sec:con}.

\section{Distribution of solar system C/O}\label{sec:CtoO}

\subsection{Stellar C/O}

The stellar C/O is set by the chemical properties of gas and dust that the star accretes during its formation. Here we outline the initial elemental ratios that are inferred from observations of star forming regions and from within our own solar system, and show how these can be interpreted in the context of planet formation.

Often, the initial chemical state of the gas disk is assumed to match the molecular abundances of dark clouds, meaning it did not undergo any chemical processing during the initial formation of the protostar-disk system - this is the so called `inheritance' scenario (featured in \cite{Fogel11,Cleeves14,Walsh15,Crid16a,Eistrup2016}). Another plausible scenario is that the gas and ice chemically evolves as it accretes onto the disk \citep{Visser2009,Visser2011,Drozd14,Drozd16}. A final scenario is known as `reset' - when all of the molecules are dissociated into their base elements \citep{Visser2009,Pon14,Eistrup2016}. While the distinction between any of the scenarios is not important to the discussion of C/O (after all, the number of carbon and oxygen atoms in the gas and ice does not change), which scenario initializes the chemical state of the gas disk can have a drastic impact on its chemical evolution of its volatiles \citep{Eistrup2016}. In this work we will follow the initial volatile elemental abundance of \cite{Eistrup2016}, with C/O $=0.34$, and assume that the volatile molecules are inherited from the pre-stellar core.

For now, we similarly assume that the initial chemical state of the refractories follows the interstellar medium (ISM), meaning that there was no chemical interaction between the gas and refractory dust during the collapse of the cloud. If the chemical composition of the dust is representative of the ISM then it has an atomic carbon-to-silicon ratio C/Si$_{\rm ISM}$ $=6$ \citep{Berg15}. Based on chemical equilibrium simulations of condensing refractories in protoplanetary disks, \cite{Alessi16} find that the most abundant silicates have silicon-to-oxygen ratios no lower than $\sim 1/4$ - mostly enstatite (MgSiO$_3$) or forsterite (MgSiO$_4$). While they produce oxygen-bearing iron species (fayalite - Fe$_2$SiO$_4$, ferrosilite - FeSiO$_3$) within 5 AU, the majority of iron-bearing material is troilite (FeS) and pure iron. When iron-oxides become more abundant, it is at the expense of enstatite, meaning that C/O remains unchanged - implying that refractories are generally carbon rich relative to oxygen (C/O $>1$). If we assume that the dust is composed of only enstatite (Si/O $=1/3$, C/O $=2$) and the Sun accretes gas and dust with the characteristic gas-to-dust ratio (100) then we can estimate the photospheric C/O. The mean molecular weight of the gas is 2.4 g mol$^{-1}$, while the molecular weight of the dust is 172.4 g mol$^{-1}$ (6 carbon atoms per enstatite molecule - m$_{\rm mol,enstatite} = 100.4$ g mol$^{-1}$). Hence:\begin{align}
m_{\rm dust} =& m_{\rm gas}/100 \nonumber\\
n_{\rm dust}\cdot m_{\rm mol,grain} =& n_{\rm H}\cdot m_{\rm mol,gas}/100 \nonumber\\
n_{\rm dust} =& \left(m_{\rm mol,gas}/m_{\rm mol,grain}/100\right) ~n_{\rm H},\nonumber
\end{align}
where $n_{dust}$ is the number of groupings of 6 carbon atoms and one enstatite molecule. So (see a summary in Table \ref{tab:chem01})\begin{align}
n_{\rm C,gr} = \frac{6}{7} n_{\rm dust}\nonumber\\
n_{\rm SiO_3,gr} = \frac{1}{7} n_{\rm dust}\nonumber
\end{align}
Hence:\begin{align}
n_{\rm C,gr} = 1.2\times 10^{-4} ~n_{\rm H} \nonumber\\
n_{\rm Si,gr} = 2.0\times 10^{-5} ~n_{\rm H}, \nonumber\\
n_{\rm O,gr} = 6.0\times 10^{-5} ~n_{\rm H}.
\label{eq:intro01}
\end{align} 

\begin{table}
\caption{Carbon-to-hydrogen (C/H) and carbon-to-oxygen ratios (C/O) for the volatiles (gas), refractories (gr), and combined (tot) given the simple model (mdl) where dust mass is dominated by enstatite and solid carbon (with C/Si = 6). Note that C/O$_{\rm tot}$ = (C$_{\rm gas}$ + C$_{\rm ref}$) / (O$_{\rm gas}$ + O$_{\rm ref}$).}
\begin{center}
\label{tab:chem01}
\begin{tabular}{|c|c|c|c|}
\hline
C/H$_{\rm gas}$ & 1.8$\times 10^{-4}$ & C/H$_{\rm gr,mdl}$ & 1.2$\times 10^{-4}$ \\\hline
C/O$_{\rm gas}$ & 0.346 & C/O$_{\rm gr,mdl}$ & 2.000 \\\hline
C/O$_{\rm tot,mdl}$ & 0.517 & C/O$_{\rm solar}$ & 0.54 \\\hline
\end{tabular}
\end{center}
\end{table}

The volatile carbon and oxygen abundances of \cite{Eistrup2016} are $n_{\rm C,gas} = 1.8\times 10^{-4} ~n_{\rm H}$, and $n_{\rm O,gas} = 5.2\times 10^{-4} ~n_{\rm H}$. Combining these values with our computed elemental abundances from our simple refractory model, we predict a solar photosphere with C/O $=0.52$, and C/Si $=15$ which is close to the observed values for the Sun (0.54, $\sim 10$, \cite{Berg15}). We can conclude two things about the solar photosphere: 1) it is carbon rich relative to the gas inherited in the protoplanetary disk, and 2) it has accreted refractories with a gas-to-dust mass ratio of at most 100\footnote{Radial drift is a known physical process that brings dust into the inner disk faster than viscous evolution. So in principle the Sun has accreted material with a lower gas-to-dust mass ratio than 100. Low gas-to-dust ratios would lead to large estimates of $n_{\rm dust}$ and hence larger elemental abundances in refractories}. 

\subsection{Implication to planet formation}

Following the above logic implies that the planets produced in \cite{Crid17} will have similarly sub-stellar C/O as in \cite{Thiabaud2015} and \cite{Mordasini16}, but for different reasons. In \cite{Thiabaud2015} and \cite{Mordasini16} the total disk (refractories \& volatiles) C/O is set to solar and the carbon/oxygen component of the planet atmosphere is predominately set by the accretion of icy planetesimals directly into the (otherwise pure H/He) atmosphere. Because ice tends to be oxygen-rich, they predict sub-stellar C/O. In \cite{Crid17}, the initial disk volatile C/O (=0.288) is dictated by the inheritance of gas from a dark cloud chemical simulation, which has sub-stellar C/O. Since the delivery of carbon and oxygen is dominated by gas accretion in that model, their planets similarly have sub-stellar C/O.

Whether a planet inherits its carbon and oxygen inventory from the gas or icy planetesimals depends on the treatment of the radial evolution of the gas. For example \cite{Thiabaud2015}, evolve the radial distribution of oxygen and carbon bearing volatiles through the disk at the radial speed of the gas:\begin{align}
v_{gas}(r) = -\frac{3}{\Sigma r^{1/2}}\frac{\partial}{\partial r}\left(\nu\Sigma r^{1/2}\right),
\label{eq:plntform01}
\end{align}
where $\Sigma$ is the surface density of the gas, and $\nu$ is the strength of viscosity.

By ignoring the transport of volatiles frozen onto grains, eq. \ref{eq:plntform01} implies that elements heavier than helium are cleared from the planet formation region of the disk (r $\lesssim 3$ AU) in less than a Myr \citep{Thiabaud2015}. Including the transport and sublimation of frozen volatiles, \cite{Bosman2017b} show that the abundance of oxygen and carbon carrying volatiles can be maintained near their initial abundances over longer timescales ($\gtrsim 1$ Myr), which would make them available for accretion onto a planetary atmosphere. Here we assume that carbon and oxygen bearing volatiles do not evolve radially, and hence only their \textit{chemical} evolution is important to the abundance of elemental carbon and oxygen in either the gas or ice phases. 

The sub-stellar C/O results discussed above are currently at odds with the previously mentioned observational work that suggests that planetary atmospheres have either stellar or super-stellar C/O \citep{Brewer16}.

\subsection{Refractory carbon depletion}\label{sec:refdepl}

A final detail regarding the distribution of carbon in the protoplanetary disk was pointed out by \cite{Mordasini16}, and is based on past work \citep{Allegre2001,Lee2010,Berg15} - namely carbon depletion. While the ISM is carbon rich relative to silicon, the Earth is depleted in carbon by more than 3 orders of magnitude \citep{Berg15}. Similarly, some asteroids show intermediate levels of carbon depletion, leading \cite{Mordasini16} to derive a simple radial distribution for the level of carbon depletion in the solar nebula:\begin{align}
{\rm C/Si}(r) = \begin{cases}
0.001 & r_{\rm AU} \le 1 \\
0.001 r_{\rm AU}^c & 1 < r_{\rm AU} < 5 \\
6 & r_{\rm AU} \ge 5 \\
\end{cases},
\label{eq:intro02}
\end{align}
where $r_{\rm AU}$ is the orbital radius in units of AU, and $c \sim 5.2$ is used to connect the outer ($r \ge$ 5 AU) and inner ($r \le$ 1 AU) parts of the disk. Because of this refractory carbon depletion, solids in the \cite{Mordasini16} model are generally more oxygen rich than assumed here deriving the abundances in eq. \ref{eq:intro01}. Such a refractory carbon depletion has been proposed to exist in other solar systems by studying the pollution of white dwarf stars, which show sub-solar carbon abundances indicative of the accretion of shredded carbon-poor planetesimals \citep{Jura2008}.

\begin{figure}
\centering
\includegraphics[width=0.5\textwidth]{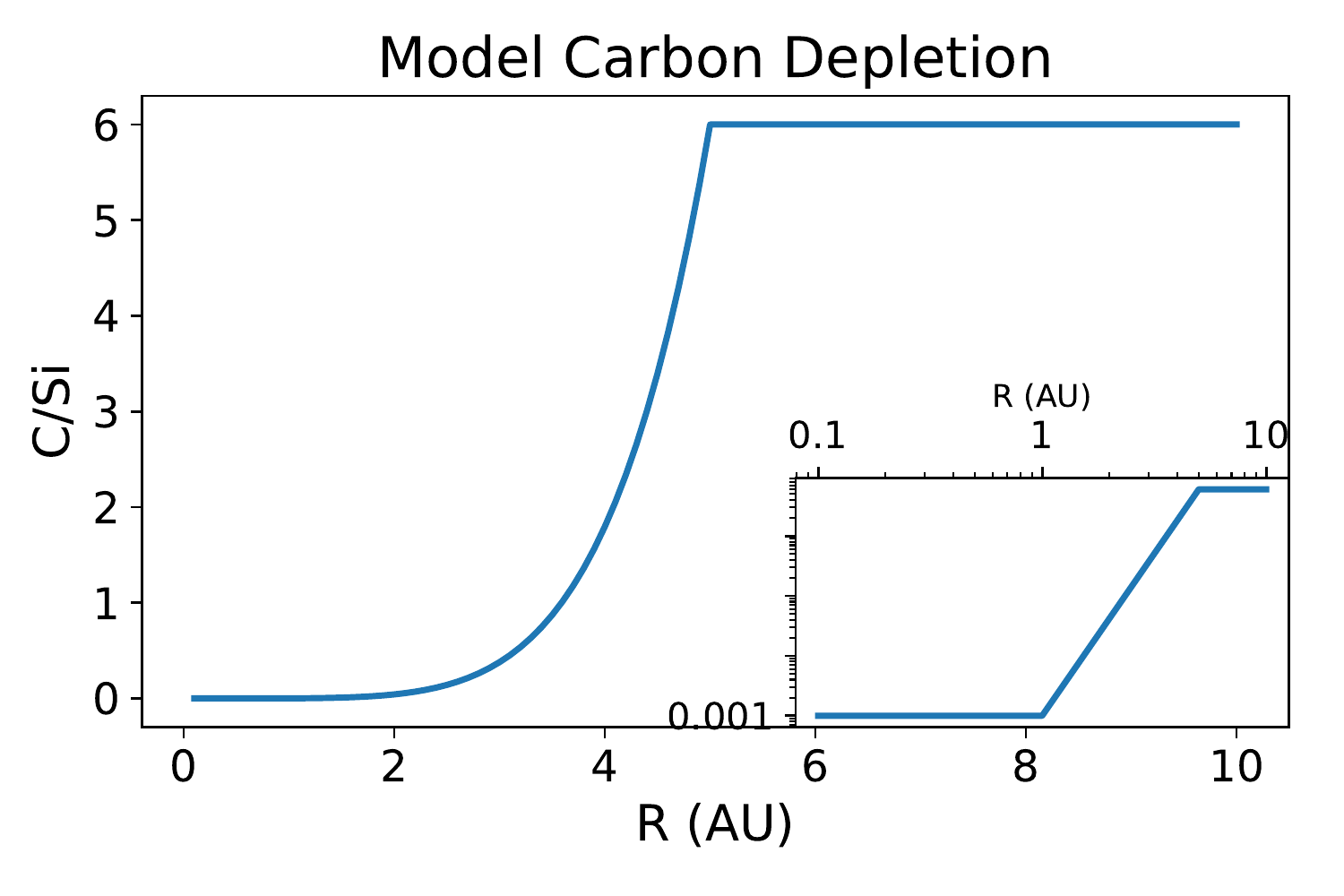}
\caption{Radial dependency of carbon depletion as described in eq. \ref{eq:intro02}. The inset shows the same plot in log-space, mimicking Figure 2 of \citet{Mordasini16}. }
\label{fig:01}
\end{figure}

In Figure \ref{fig:01} we show the graphical form of eq. \ref{eq:intro02} in linear and log-space (inset). The source of this carbon depletion is not well constrained (see for ex. \cite{Berg15}). Two plausible explanations involve ongoing processing of the refractory carbon through oxidation and photodissociation \citep{Lee2010,Anderson2017}, or early thermal processing of the grains as they are accreted onto the disk \citep{Berg15} - similar to the reset scenario discussed earlier. In both cases the carbon is returned to the gaseous state, and would be available for accretion onto a planetary atmosphere.

The carbon that is returned to the gas quickly reacts with water to create CO. Within 1 AU, the maximum amount of carbon that is released into the gas is nearly all of the solid carbon budget inherited from the ISM. In the ISM, \cite{MishraLi2015} report Si/H$_{\rm ISM}$ $= 40.7$ ppm which is twice as high as the value that we derive in the simple model above. When combined with C/Si$_{\rm ISM}=6$, we find an excess carbon abundance of C/H$_{\rm exc} = 2.44\times 10^{-4}$ (i.e. 2/3 of the total carbon) that would be released into the gas phase. This excess gaseous carbon leads to a gas disk C/O = 0.82 inside the water ice line. 

Along with the increase in the refractory carbon in the disk we similarly expect a higher available refractory oxygen. To estimate this refractory component we follow \cite{Mordasini16} who assume that the refractories have a mass ratio of 2:4:3 for the carbon, silicates, and irons respectively. This assumption leads to a refractory oxygen abundance of O/H$_{\rm ref} = 1.75\times 10^{-4}$. 

\ignore{
\begin{figure}
\centering
\subfloat[Available excess carbon for atmospheric accretion. Here assume that some ongoing process is supplying the excess carbon to the midplane gas.]{
\includegraphics[width=0.5\textwidth]{excess_carbon_ongoing.pdf}
\label{fig:02a}
}\\
\subfloat[Same as Figure \ref{fig:02a} but assuming that the carbon excess was part of the initial elemental abundance of the gas, due to a reset scenario. In that case the excess carbon will advect with the gas as it accretes onto the host star. ]{
\includegraphics[width=0.5\textwidth]{excess_carbon.pdf}
\label{fig:02b}
}
\caption{Two possible carbon excess models.}
\label{fig:02}
\end{figure}
}

\begin{figure}
\centering
\includegraphics[width=0.5\textwidth]{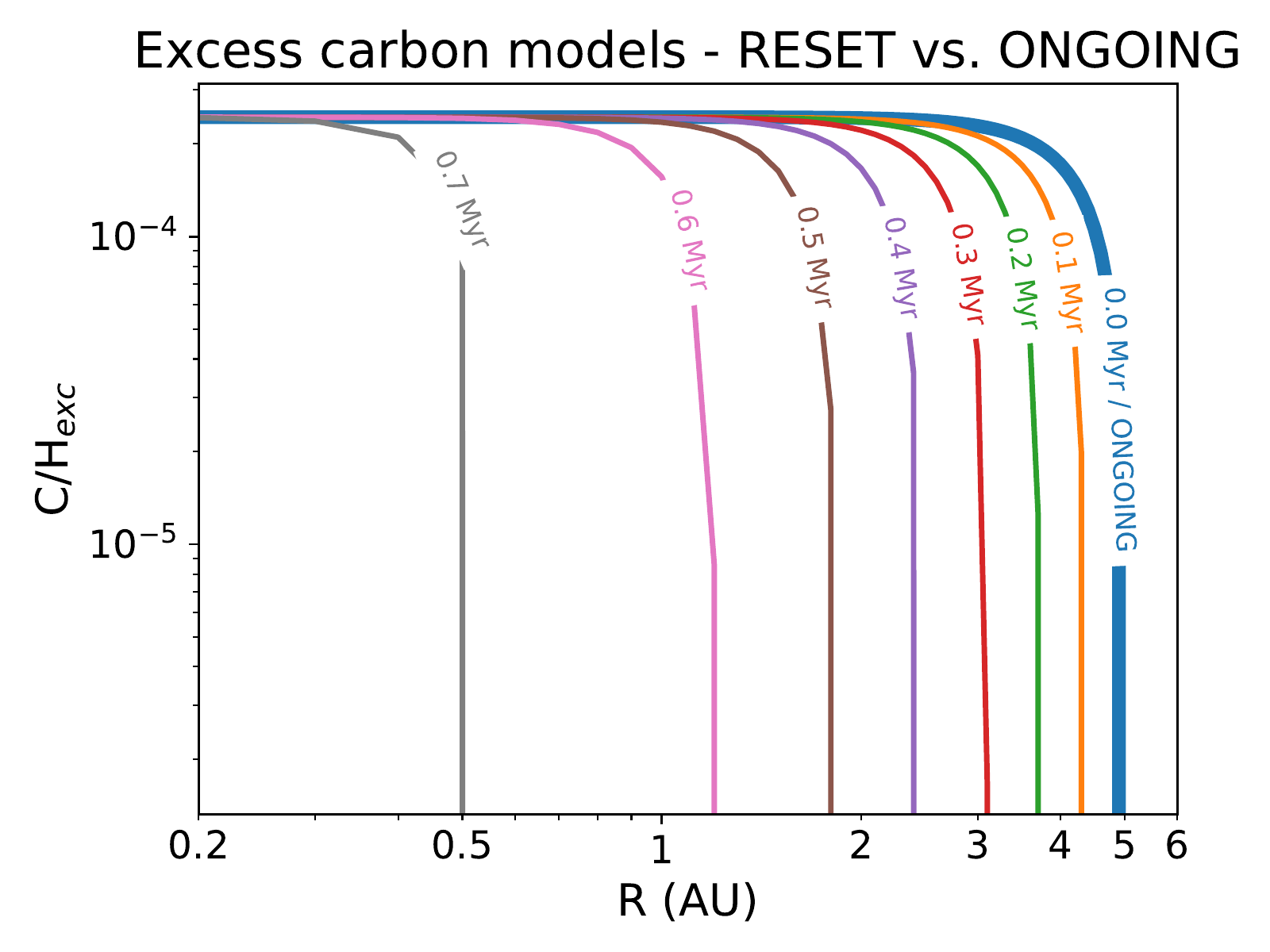}
\caption{Available excess carbon for atmospheric accretion. Here we assume that in the `ongoing' model the excess carbon is supplied by ongoing processing of dust grains, delivering the carbon to the gas. In the `reset' model the carbon excess is generated early in the disk life, and advects with the bulk gas into the host star.}
\label{fig:02}
\end{figure}

In Figure \ref{fig:02} we show two toy models for the amount of excess carbon that is available in the gas, given the two depletion mechanisms mentioned above. In the `ongoing' scenario oxidation or photochemistry maintains the high abundance of gaseous carbon shown in Figure \ref{fig:01} over the lifetime of the gas disk. In this model, while the gas is advected into the host star rapidly (see below), inflowing dust particles from larger radii (through radial drift) are processed efficiently to maintain the carbon excess in the gas that is inferred by the carbon depletion observed today in the solids on Earth. Hence we assume that the carbon excess is maintained for the whole lifetime of the disk.

In the `reset' scenario we assume that all of the excess carbon is delivered early in the disk's lifetime. This excess would then advect with the gas as it accretes onto the host star. Because there is no excess carbon advecting from larger radii, its carbon excess is not regenerated as in the `ongoing' scenario. We model this process by evolving the shape of the `ongoing' model (blue curve in Figure \ref{fig:02}) inward with an assumed constant advection speed $v_{\rm adv}$. The time evolving version of eq. \ref{eq:intro02} becomes\begin{align}
{\rm C/Si}(r,t) = \begin{cases}
0.001 & r_{\rm AU} \le 1 - v_{\rm adv}t \\
0.001 r_{\rm AU}^c & 1 - v_{\rm adv}t < r_{\rm AU} < 5 - v_{\rm adv}t \\
6 & r_{\rm AU} \ge 5 - v_{\rm adv}t \\
\end{cases},
\label{eq:intro03}
\end{align}
such that the excess carbon in the gas phase is:\begin{align}
{\rm C/H}_{\rm exc} = ({\rm C/Si}_{\rm ISM} - {\rm C/Si}(r,t))\times {\rm Si/H}_{\rm ISM},\nonumber
\end{align}
where we take $v_{\rm adv} = 0.03$ m s$^{-1}$ $=6.3$ AU Myr$^{-1}$, which represents a typical advection speed from \cite{Bosman2017b} for a gaseous disk with viscous $\alpha=10^{-3}$. The time evolution shown in Figure \ref{fig:02} is qualitatively similar to the evolution of molecular gas species shown in \cite{Thiabaud2015}. As in that work, we find the radial extent of the excess carbon in the gas is lost to the host star by 0.8 Myr into the lifetime of the disk. 

We note here that carbon depletion does not affect the earlier prediction of the stellar photosphere, since the excess carbon that is processed off of the grains would be delivered to the gas and accrete towards the host star. Hence the number of carbon atoms accreting through the disk is preserved.

\begin{figure*}
\centering
\begin{overpic}[width=\textwidth]{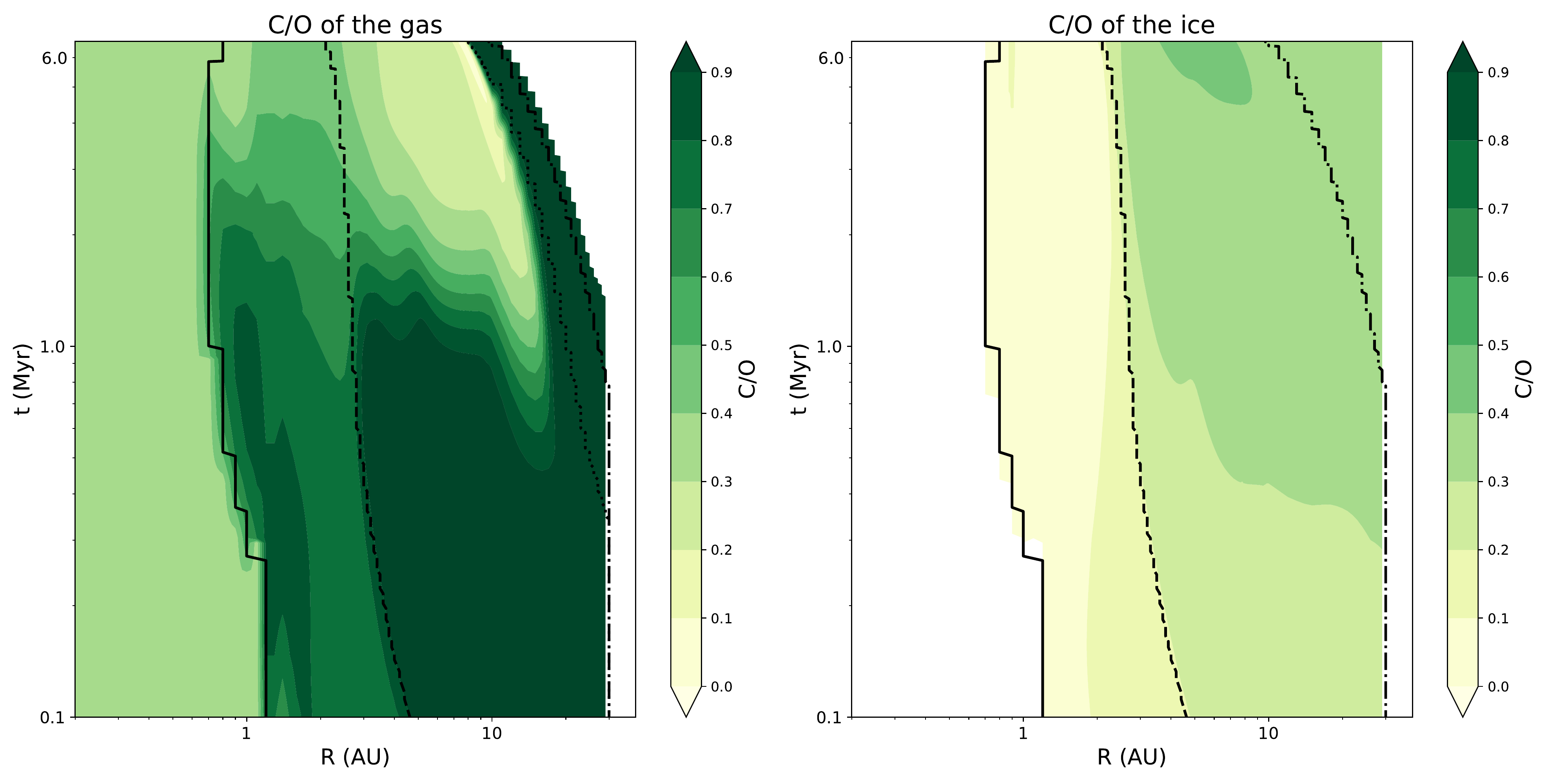}
\put(9,40){H$_2$O}
\put(22,40){CO$_2$}
\put(30,40){O$_2$}
\put(36,40){CO}
\end{overpic}
\caption{The evolution of C/O in gas (left) and ice (right) from \citet{Eistrup2017}. We note the location of the water (solid), CO$_2$ (dashed), O$_2$ (dotted) and CO (dot-dashed) ice lines. A step-function in C/O is seen early on in the disk's life which becomes complicated by chemistry as the disk ages. After $\sim $ Myr CO is converted into frozen CO$_2$ outward of the CO$_2$ ice line, which lowers the gas C/O while raising the ice C/O - eventually becoming more carbon rich than the gas.}
\label{fig:chem01}
\end{figure*}

\section{ Disk / Chemical Model }\label{sec:chemical}

For the gas and ice abundances in the disk we use the chemical results produced by \cite{Eistrup2017}. A benefit in using their chemical model over that presented in \cite{Crid17} is that their treatment of grain-surface chemistry is more comprehensive, and includes a more extended network of chemical reactions that are relevant for determining the final abundance of carbon and oxygen-bearing molecules in both the gas and ice phases. 

The most important result of \cite{Eistrup2017} is the evolution of the C/O in the gas and ice. Because of their network of surface reactions, \cite{Eistrup2017} found a departure from the typical `step function' description of C/O as first pointed out by \cite{Oberg11}. In particular, through the destructive reaction of CO in gas-grain interactions:\begin{align}
{\rm iCO + iOH \rightarrow iCO_2 + iH}\nonumber
\end{align}
between the CO$_2$ and O$_2$ ice lines, carbon and oxygen are slowly converted from the gas phase to the ice phase (denoted by the lower case `i'). In a similar region of the disk (inward of the O$_2$ ice line) molecular oxygen can similarly be produced on the grain surface through the reaction \citep{Eistrup2017}:\begin{align}
{\rm iO + iOH \rightarrow iO_2 + iH},\nonumber
\end{align}
and subsequent desorption of O$_2$.

The disk model used in \cite{Eistrup2017} was based on a power-law fit to numeric results of \cite{Alibert2013}. Key features of the disk model include its cooling gas temperature and lowering surface density as the disk ages. This evolution is the main driver of the inward motion of the volatile ice lines, but also impacts the flux of ionization radiation, which depends on the surface density of the gas. The main driver of gas ionization along the disk midplane is cosmic-ray ionization whose rate depends exponentially on the gas surface density (see \cite{Eistrup2017} for details).

In Figure \ref{fig:chem01} their evolution of C/O is presented along with the locations of the water (solid), CO$_2$ (dashed), O$_2$ (dotted) and CO (dot-dashed) ice lines.  The step-function in C/O with steps at each of the ice lines is clearly seen, and they slowly move inward as the disk ages and cools. Outward of the CO$_2$ ice line at times later than $\sim$ Myr we see the conversion of gaseous CO into frozen CO$_2$ which enhances the frozen C/O at the expense of the gaseous C/O.

This enhancement occurs between the CO$_2$ and O$_2$ ice lines where the most dominant molecular species in the gas (heavier than He) are CO and O$_2$ (because of its production mentioned above), hence as CO freezes out, the gas becomes more oxygen rich (C/O decreases). The icy volatile component begins oxygen rich (primarily CO$_2$ and H$_2$O), and adding equal amounts of carbon and oxygen (ie. by freezing out CO) will generally increases C/O. This exchange occurs inward of the O$_2$ ice line, where CO should tend to be gas. However, outside of the CO$_2$ ice line when a CO molecule happens to stick to a dust grain its reaction with frozen OH proceeds faster than its desorption back to the gas phase.

This exchange of elements between the ice and gas could be important to the chemical composition of planetary atmospheres because of the relative amount of solid and gas accretion that is responsible for enhancing the atmosphere in heavier (than helium) elements.

\section{ Planet formation }\label{sec:plntform}

Currently the process of planet formation through core accretion (in oppose to planet formation through gravitational instability which is not the subject of this paper) is split into two dominant methods: planetesimal and pebble accretion. The former is discussed below, and involves the accretion of large (10-100 km) bodies, while the latter describes growth through the accretion of mm-cm sized `pebbles' \citep{Ormel2010,LambJoh2014,Bitsch2015}.  A particular advantage of accreting pebbles over planetesimals is that the process is at least a few orders of magnitudes faster \citep{Bitsch2015} which offers a second way of circumventing the previously mentioned `Type-I migration problem'.

Along with their differences in accretion rates, pebbles may also alter the delivery of heavy elements to the atmosphere of forming planets because their survivability through a proto-atmosphere is lower than that of planetesimals \citep{Mordasini2006,Brouwers2018}, and hence they should tend to deliver more refractory material to the upper atmosphere (for ex. as studied by \citealt{AliDib2017}) than planetesimals. While this offers a possible method of differentiating the two formation mechanisms, combining pebble accretion with our chemical model is currently beyond the scope of this work.

The planet formation model used here is outlined in \cite{Crid17} and assumes the standard planetesimal accretion paradigm. Here we outline some key features of the model.

\subsection{Solid accretion}\label{sec:solidacc}

In the planetesimal accretion model of planet formation (ie. \cite{Pollack1996}, \cite{KI02}, \cite{IL04a}) planetary cores grow by first accreting from a disk of 10-100 km sized planetesimals. Known as the oligarchic phase, it has an accretion timescale of \citep{KI02,IL04a}: \begin{align}
t_{c,acc} &= 1.2\times 10^5 {\rm yr} \left(\frac{\Sigma_d}{10 {\rm g~cm}^{-2}}\right)^{-1}\left(\frac{a_p}{\rm AU}\right)^{1/2}\left(\frac{M_c}{M_\oplus}\right)^{1/3}\nonumber\\
&\times \left(\frac{M_*}{M_\odot}\right)^{-1/6}\left[\left(\frac{\Sigma_g}{2.4\times 10^3 {\rm g~cm}^{-2}}\right)^{-1/5}\right.\nonumber\\
&\times \left. \left(\frac{a_p}{\rm AU}\right)^{1/20}\left(\frac{m}{10^{18}{\rm g}}\right)^{1/15}\right]^2,
\label{eq:solidacc}
\end{align}
which includes the effect of gas drag on the incoming planetesimals (terms in the square brackets). The surface density of solids ($\Sigma_d$) and gas ($\Sigma_g$) are evaluated at the radial position of the planet ($a_p$) at each timestep, and the incoming material is added to the mass of the planetary core ($M_c$) through: $M_c(t+dt) = M_c(t)(1 + dt/t_{c,acc})$. 

Globally we assume a gas-to-dust ratio of 100, such that $\Sigma_d = 0.01\Sigma_g$, however as has been assumed in the past (ie. see \citealt{Alessi16}) we enhance the amount of solid material in the planet-forming region by an order of magnitude to account for the dust trapping expected from the streaming instability (ie. \cite{Rae15}) which leads to the generation of planetesimals \citep{Schafer2017}. So while the planet is undergoing oligarchic growth, $\Sigma_d = 0.1\Sigma_g$.

We assume that the refractory component of every planetesimal is made of the same chemical composition. The carbon and oxyten content has the same elemental abundance as derived above: C/H = $2.44\times 10^{-4}$ and O/H = 1.75$\times 10^{-4}$. Meanwhile the volatile ice component follows from the results of our chemical model.

During the initial formation of the core we assume that none of the carbon and oxygen that accretes ends up in the atmosphere. This reasoning comes from work by \cite{Schlichting2015} who have shown that planetesimal accretion events are efficient at eroding atmospheres, particularly around low mass planets. This implies that any volatiles that might be out-gassed during the initial formation of the core are unlikely to survive as a proto-atmosphere. So in accounting for the elements in the atmosphere we will ignore the contribution of the material that originally formed the core. 

At a certain point in its evolution, the growing planet consumes or scatters most of the planetesimals in its immediate vicinity. At this point the heating from impacting planetesimals is no longer sufficient to maintain pressure support of the surrounding gas, and the planet transitions from a stage of oligarchic growth to a stage of gas accretion. When this happens we assume that the original population of planetesimals is no longer present, and hence we return the gas-to-dust ratio to 100. We continue to allow planetesimal accretion during the gas accretion phase to occur (now with $\Sigma_d = 0.01\Sigma_g$), and it is at this point that the delivery of heavy elements to the atmosphere can begin. 

As the gas envelope begins to develop we start counting the elements that are delivered into the atmosphere by solid bodies. The delivery of carbon and oxygen depends on the survival of planetesimals as they pass through the atmosphere. The thermal ablation and aerodynamic fragmentation of a planetesimal has been calculated in detail by \cite{Mordasini15}. Here we take a simple approach and assume that if the atmosphere is lighter than 3 M$_\oplus$ the planetesimal can fully penetrate and reach the core, while above 3 M$_\oplus$ the planetesimal is completely destroyed in the atmosphere. In the former case, only the most volatile species (ex. CO$_2$, H$_2$O) are released into the atmosphere, if they are frozen to the incoming planetesimals as computed by the chemical model; while in the latter case we assume that both the volatile ices and the refractories are distributed into the atmosphere. The critical atmosphere mass was chosen by inspecting figure 9 in \cite{Mordasini15} and selecting the highest atmospheric mass where planetesimals of all sizes are destroyed in the atmosphere. This is a necessary simplification since we compute neither the internal structure of the atmosphere, nor the size distribution of the incoming planetesimals.

We note that in computing the carbon and oxygen content of the icy planetesimals we have generally ignored their radial transport that could enhance the local gas C/O (as in \cite{Booth2017}) or the atmosphere directly by volatile delivery through N-body interactions after the gas disk has evaporated.

Later in its evolution, the planet opens a gap in the gas disk (see below). The gap stops the flow of dust through the disk because it forms a positive pressure gradient which traps most of the dust that is decoupled from the gas \citep{Pinilla2016,Bitsch2018}. However the smallest grains which are still coupled to the gas dynamics can still flow accross the gap \citep{Bitsch2018}, but these grains are generally about two orders of magnitude less abundant (by mass) than the typical mm-cm sized grains \citep{Crid16b}. Additionally, the gas drag which contributes to widening the proto-planet's effective feeding zone (square brackets in eq. \ref{eq:solidacc}) vanishes. Without the gas drag, planetesimal accretion rates are reduced by about two orders of magnitude \citep{IL04a}. Hence we expect the rate of solid accretion will drop when the planet opens a gap. To account for the change we reduce the surface density of solids available for accretion by two orders of magnitude ($\Sigma_d = 0.0001\Sigma_g$) when the gap is opened.

Finally we set a global maximum to the amount of solid material that can be accreted by a planet. This global maximum is defined as the total solid mass in the disk, which we estimate by integrating the gas surface density profile over the whole disk (radii between 0.01 AU and 30 AU) and scaling it by a gas-to-dust ratio of 100. 

\subsection{Gas accretion}

When the planet reaches a critical mass it can begin to draw down the gas from the surrounding protoplanetary disk. This process is limited by the rate of heating that the planet receives from infalling planetesimals, and the rate at which the gas can radiate away its gravitational energy. Both of these processes ultimately depend on the opacity of the envelope ($\kappa_{env}$) \citep{Ikoma2000,IdaLin2008,HP14,Mordasini14b,Alessi16b}, with the critical mass having the form:\begin{align}
M_{c,crit} = 10 M_\oplus\left(\frac{1}{10^{-6}M_\oplus yr^{-1}}\frac{dM_p}{dt}\right)^{1/4}\left(\frac{\kappa_{env}}{1 ~{\rm cm^2g^{-1}}}\right)^{0.3},
\label{eq:gas01}
\end{align}
and the mass accretion rate depends on the Kelvin-Helmholtz timescale:\begin{align}
t_{KH} = 10^c{\rm yr}\left(\frac{M_p}{M_\oplus}\right)^{-d},
\label{eq:gas02}
\end{align}
where $c$ and $d$ depend on $\kappa_{env}$ (see \cite{Alessi16b} for their description). Given this timescale the rate at which gas is accreted onto the planet is:\begin{align}
\dot{M_p} = M_p/t_{KH}.
\label{eq:gas03}
\end{align}
Given their population synthesis model, \cite{Alessi16b} pick a nominal opacity of $\kappa_{env} = 0.001 ~{\rm cm^2g^{-1}}$ which best reproduces the population of observed exoplanets - with this opacity: $c=7.7$, and $d=2$. 

Eventually the planet has grown to a mass of $\sim 10$ M$_\oplus$, at which point it opens a gap in the protoplanetary disk. When this happens the planet decouples from the surrounding disk and the geometry of the gas accretion changes \citep{Szul2014}. In this regime the gas primarily accretes vertically onto the planet from above the midplane of the disk. Some of the gas is circulated back into the surrounding disk as described by \cite{Morbidelli2014} and illustrated by \cite{Batygin2018}. 

As described by \cite{Batygin2018}, as the gas falls towards the midplane it interacts with a planetary magnetic field which dictates the dynamics of the gas within the planetary magnetosphere. Within the magnetosphere gas that falls planet-ward of the critical magnetic field line crest will fall onto the planet, while gas that falls on the far side of the crest falls onto the circumplanetary disk. This gas is then recycled back into the protoplanetary disk as outlined by \cite{Morbidelli2014}. Recently, \cite{Crid2018} derived a planetary mass accretion rate based on this physical picture: \begin{align}
\frac{\dot{M}_{p,Mag}}{M_\oplus/yr} = \frac{4}{3^{3/4}}\left(\frac{R_0}{R_H}\right)^2 \left(\frac{M_p}{M_\oplus}\right)^{-2/7}\left(\frac{\dot{M}}{M_\oplus/yr}\right)^{3/7},
\label{eq:math03}
\end{align}
where $\dot{M}$ is the gas accretion rate into the gap, $R_H = a_p (M_p/3M_*)^{1/3}$ is the Hill radius, $R_0 = (\pi^2/2\mu_0 ~\mathcal{M}^4/GM_\oplus^2/yr)^{1/7}$ is a scaling factor with units of length, and $\mathcal{M}=B R_p^3$ is the magnetic moment for an (assumed) magnetic dipole, of magnetic field strength $B$ and planetary radius $R_p$.

As argued in \cite{Crid2018}, we set $B=500$ Gauss for the young planet, which assumes that the field is in equipartition with the kinetic energy of the convecting part of the planetary interior. This strength is about two orders of magnitude stronger than Jupiter's magnetic field today, but weaker by a factor of a few than a typical brown dwarf. In effect, setting the field strength in this way represents an interpolation between our understanding of the geo- and solar- dynamo \citep{Christensen2009}. For simplicity we keep the magnetic field strength constant throughout the growth of the planet, assuming that the majority of the gravitational energy of infalling gas is radiated away (ie. a cold start model). Generally we expect the mass accretion rate to weakly depend on the magnetic field strength - since $\dot{M}_{p,mag} \propto B^{8/7}$.

As shown in \cite{Mordasini13}, the size of the planet stays nearly constant after it has opened a gap in its disk. Hence we similarly assume that $R_p$ does not change during this phase of planet formation. We assume that the planet has undergone a `cold-start', such that the planet has not stored a large amount of entropy during its early stage of gas accretion (see \cite{Mordasini13}), and $R_p=1.5 R_J$.

When the gap is opened, the gas accretion is still limited by the rate at which it can release its gravitational energy, hence the true mass accretion rate is the smaller of eqs. \ref{eq:gas03} and \ref{eq:math03}:\begin{align}
\dot{M}_p = {\rm min}\left(M_p/t_{KH},\dot{M}_{p,Mag}\right).
\end{align}

We place a similar maximum gas mass on the gas accretion as we did for the solid accretion, since our disk is on the lower mass end of known disk masses with M$_0 \sim 5$ M$_J$ ($\sim 0.005$ M$_{\odot}$) of gas. Hence when the planet exceeds this gas mass, it is reverted to its mass of the previous time step. 

\subsection{Planet Migration}

The planet interacts gravitationally with the gas in its protoplanetary disk, which exchanges angular momentum. The result of this interaction is that the planet changes its orbital radius, a process often called `planet migration'. The rate of migration depends on the radial structure of the gas temperature and density structure. Recently Cridland et al. (submitted) outlined a method of computing the rate of planetary migration for a given disk model. The method is based on the work of \cite{Paard11} and \cite{Cole14} and includes the effect of transitions in the dust opacity and turbulent $\alpha$ at ice lines and the dead zone edge respectively.

Here we follow the methods of Cridland et al. (submitted) and compute a modified temperature profile based on a combination of the profile used in \cite{Eistrup2017}, and a radially dependent dust opacity which depends on the local abundance of water and CO$_2$ ice. Generally this warms the gas within the water ice line, because the dust opacity is higher when it is not covered with a layer of ice, and flattens the temperature profile over the remainder of the disk. At the original ice line location of the \cite{Eistrup2017} disk model the gas is warmed by about 50\% so that in the new disk model the ice line is shifted out by about 0.5 AU. In principle this shift changes the distribution of volatile C/O in the new disk model however such a small change will not drastically impact the chemical signature of the disk on the chemical composition of planetary atmospheres.

Changing the temperature profile has a larger effect on the total torque on the planet, which has the form \citep{Cole14}:\begin{align}
\Gamma_{I,tot} &= \Gamma_{LR} + \left[\Gamma_{VHS}F_{p_\nu}G_{p_\nu} + \Gamma_{EHS}F_{p_\nu}F_{p_\chi}\sqrt{G_{p_\nu}G_{p_\chi}} \right. \nonumber\\
&+ \left. \Gamma_{LVCT}\left(1-K_{p_\nu}\right) + \Gamma_{LECT}\sqrt{\left(1-K_{p_\nu}\right)\left(1-K_{p_\chi}\right)}\right],
\label{eq:torq01}
\end{align}
where $\Gamma_{LR}$, $\Gamma_{VHS}$, $\Gamma_{EHS}$, $\Gamma_{LVCT}$, and $\Gamma_{LECT}$ are the Lindblad torques, vorticity and entropy-related horseshoe drag torques, and linear vorticity and entropy-related corotation torques, respectively (see equations 3-7 in \citealt{Paard11}). The damping functions $F_{p_\nu}$, $F_{p_\chi}$, $G_{p_\nu}$, $G_{p_\chi}$, $K_{p_\nu}$, and $K_{p_\chi}$ are related to the ratio of either the viscous or thermal diffusion timescales with the horseshoe libration or horseshoe U-turn timescales (see equations 23, 30, and 31 in \citealt{Paard11}). Due to these torques the planet's angular momentum evolves such that $dL/dt = \Gamma_{I,tot}$, and:\begin{align}
\frac{da_p}{dt} = \frac{2}{M_p} \sqrt{\frac{a_p}{GM_*}}\Gamma_{I,tot},
\label{eq:torq02}
\end{align}
assuming the planet is a point mass in a circular, Keplerian orbit.

Each torque depends on the local gradient of the gas temperature and surface density. As an example, changes to the dust opacity can lead to a local change in the temperature gradient at the water ice line. The water ice line is the smallest radius in the disk midplane where water vapour is sufficiently cold to freeze onto dust grains. Within the ice line, the abundance of water ice smoothly changes from zero to its maximum abundance over a $\sim 0.5$ AU wide region. This abundance gradient leads to a smooth transition in the dust opacity because it depends on the abundance of ice frozen out on the grains (see for ex. \citealt{Miyake1993}). If the water ice line inhabits a portion of the disk that is heated primarily by viscous heating, the opacity gradient in the water ice line leads to a local flattening of the gas temperature gradient. Accordingly, to maintain a constant mass accretion rate at the water ice line, the gas surface density gradient also flattens.

These changes in gradients often result in a process known as `planet trapping' \citep{Lyra2010,Horn2012,HP11,D14}, generating points of zero torque towards which planets migrate (see details outlined in \citealt{Crid16a}). Generally it is assumed that trapped planets migrate with their trap until they reach their gap opening mass which saves computing time by eliminating the need to compute the net torque on the planet (eq. \ref{eq:torq01}). However in this work we fully compute the net torque according to eq. \ref{eq:torq01} and the resulting change in orbital radius according to eq. \ref{eq:torq02}.

Once above the gap opening mass \citep{MP06}:\begin{align}
M_{gap} = M_* {\rm min}\left(3h^3,\sqrt{40\alpha h^5}\right),
\end{align}
the torques responsible for the first type of planet migration (often called Type-I) disappear and the planet enters into a second phase of migration (Type-II). In this second phase planet migration proceeds on the viscous time: $t_\nu \sim \nu/a_p^2$, where $\nu = \alpha c_s H$ is the viscosity in the standard $\alpha$-disk model \citep{SS73} and $c_s$ and $H$ is the sound speed and scale height of the gas. With this prescription the migration rate is:\begin{align}
\frac{1}{a_p}\frac{da_p}{dt} = \frac{\nu}{a_p^2}.
\label{eq:torq03}
\end{align}
In the event that the planet becomes more massive than the total mass of the gas disk within its orbital radius (denoted by M$_{crit}$) then the rate of Type-II migration is reduced by a factor of M$_p$/M$_{crit}$.

In what follows, we will not invoke planet trapping, but instead compute the torques using eqs. \ref{eq:torq01} (using the expressions from \cite{Paard11}) and \ref{eq:torq02} until the gap is opened, then use eq. \ref{eq:torq03}.

\section{ Results: Connecting planet formation and Astrochemistry }\label{sec:results}

Here we model the formation of 11 planets, each with different initial orbital radii equally spaced between 1 and 3 AU. For this disk model, planets starting within 1 AU tend to migrate into the host star, while planets which started at radii outside of 3 AU tend not to grow very large. All planets were initialized with masses of 0.01 M$_\oplus$ and allowed to freely migrate according to the torques computed with eq. \ref{eq:torq01}. We note that while these planets will often be shown as forming in unison, in reality they are grown in isolation, and no gravitational interaction between the embryos are allowed to occur. 

\subsection{ Planet formation tracks } \label{sec:tracks}

A convenient way of discussing a planet's formation is through its formation track: the planet's evolution through the mass-semi-major axis diagram as it grows. The specific details depend on the physics included in the planet formation, migration, and disk model. Hence the cases presented here are meant as representations of what can happen.

\begin{figure}
\centering
\includegraphics[width=0.5\textwidth]{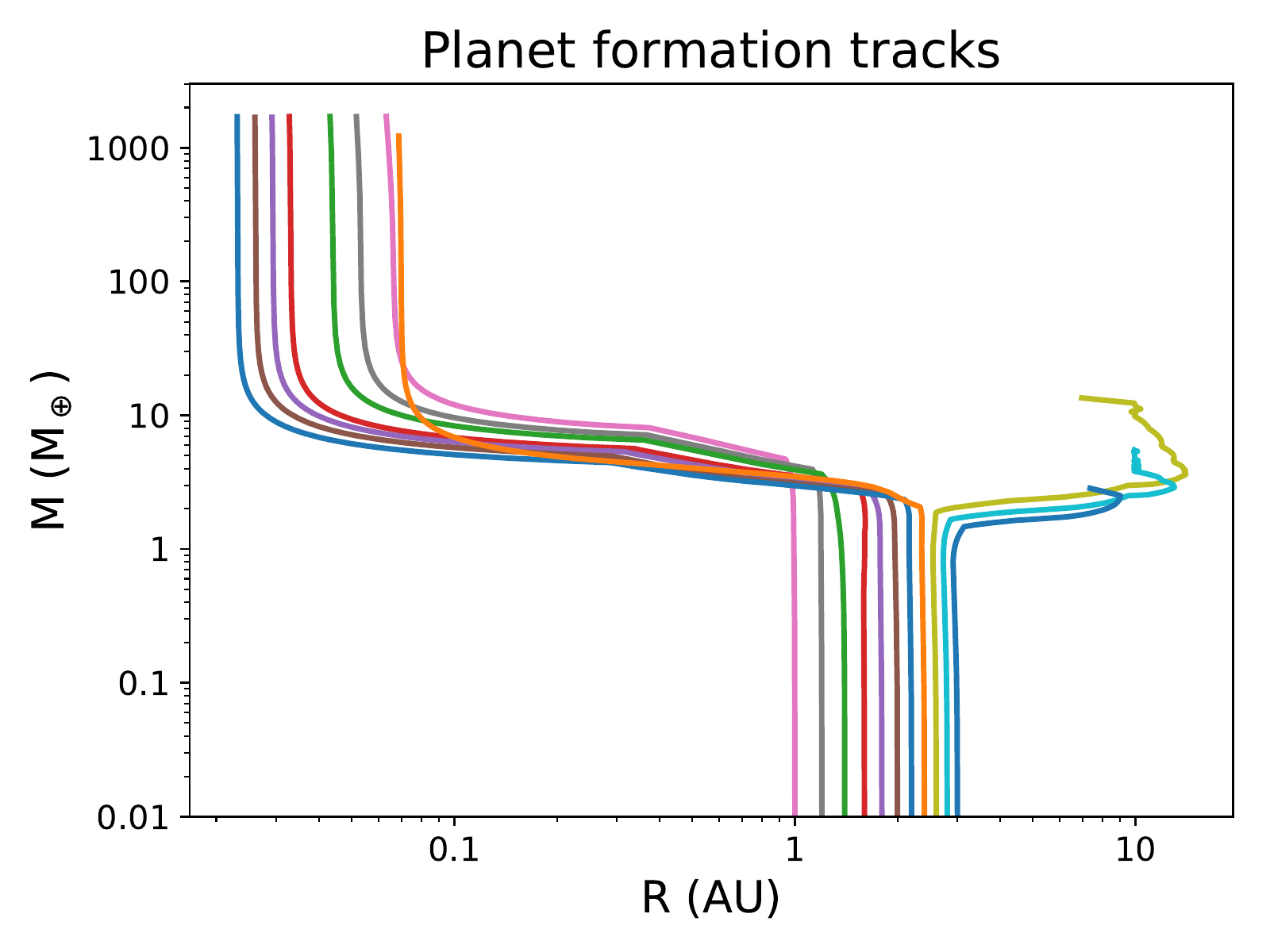}
\caption{Planet formation tracks for the synthetic planets formed in this work. There is a easily distinguishable bifurcation between the orbital evolution of the planets, caused by their interaction with the disk. }
\label{fig:tracks01}
\end{figure}

In Figure \ref{fig:tracks01} we show the planet formation tracks for the planets formed for this work. Planets that started with $a_p \le 2.4$ AU end up as Hot Jupiters, while planets starting farther out stay in the smaller super-Earth / Neptune mass range at radii between 3-10 AU. These planets' outward migration is due to their interaction with a strong positive torque (see below) caused in part by the CO$_2$ ice line, and partly due to the properties of the outer parts of our disk model (discussed below). 

The orbital properties of the planets are shown in Table \ref{tab:res01}. The general trend is that planets that start at smaller radii accrete faster, and hence do not migrate as far before their Type-II migration is slowed by their inertia. They also tend to be more massive than planets which start farther out because they accrete most of their gas in higher surface density regions of the disk. 

This trend is broken for the planet which starts at 2.4 AU. For that planet we see a very slight outward migration as it interacts with the disk. This short stage of outward migration stalls its growth enough that it only accretes $\sim$ 4 M$_J$ of gas and ends farther out than the other Hot Jupiters. Starting the planet outside of 2.4 AU places the planet in a regime of strong outward migration, moving it to lower density regions of the disk. This severely suppresses further planetary growth, resulting in the range of smaller final masses.

\begin{table}
\caption{Orbital properties of the simulated planets. }
\label{tab:res01}
\begin{tabular}{ |c|c|c| }
\hline
Initial radius (AU) & Final radius (AU) & Final Mass (M$_\oplus$) \\\hline
 1.0 & 0.06 & 1714 \\\hline
 1.2 & 0.05 & 1709 \\\hline
 1.4 & 0.04 & 1711 \\\hline
 1.6 & 0.03 & 1710 \\\hline
 1.8 & 0.03 & 1691 \\\hline
 2.0 & 0.025 & 1687 \\\hline
 2.2 & 0.022 & 1709 \\\hline
 2.4 & 0.069 & 1226 \\\hline
 2.6 & 7.0 & 13.5 \\\hline
 2.8 & 9.9 & 5.50 \\\hline
 3.0 & 7.4 & 2.85 \\\hline
\end{tabular}
\end{table}

For a more in depth study of formation history we will focus on three planets, starting at 2.2 (blue), 2.4 (orange), and 2.6 (yellow) AU.

\begin{figure*}
\centering
\ignore{
\subfloat[]{
\begin{overpic}[width=0.5\textwidth]{pic020_new.png}
\put(65,65){0.1 Myr}
\end{overpic}
\label{fig:tracks02a}
}
\subfloat[]{
\begin{overpic}[width=0.5\textwidth]{pic070_new.png}
\put(65,65){0.5 Myr}
\end{overpic}
\label{fig:tracks02b}
}\\
\subfloat[]{
\begin{overpic}[width=0.5\textwidth]{pic130_new.png}
\put(65,65){1.7 Myr}
\end{overpic}
\label{fig:tracks02c}
}
\subfloat[]{
\begin{overpic}[width=0.5\textwidth]{pic270_new.png}
\put(65,65){4.5 Myr}
\end{overpic}
\label{fig:tracks02d}
}
}
\begin{overpic}[width=\textwidth]{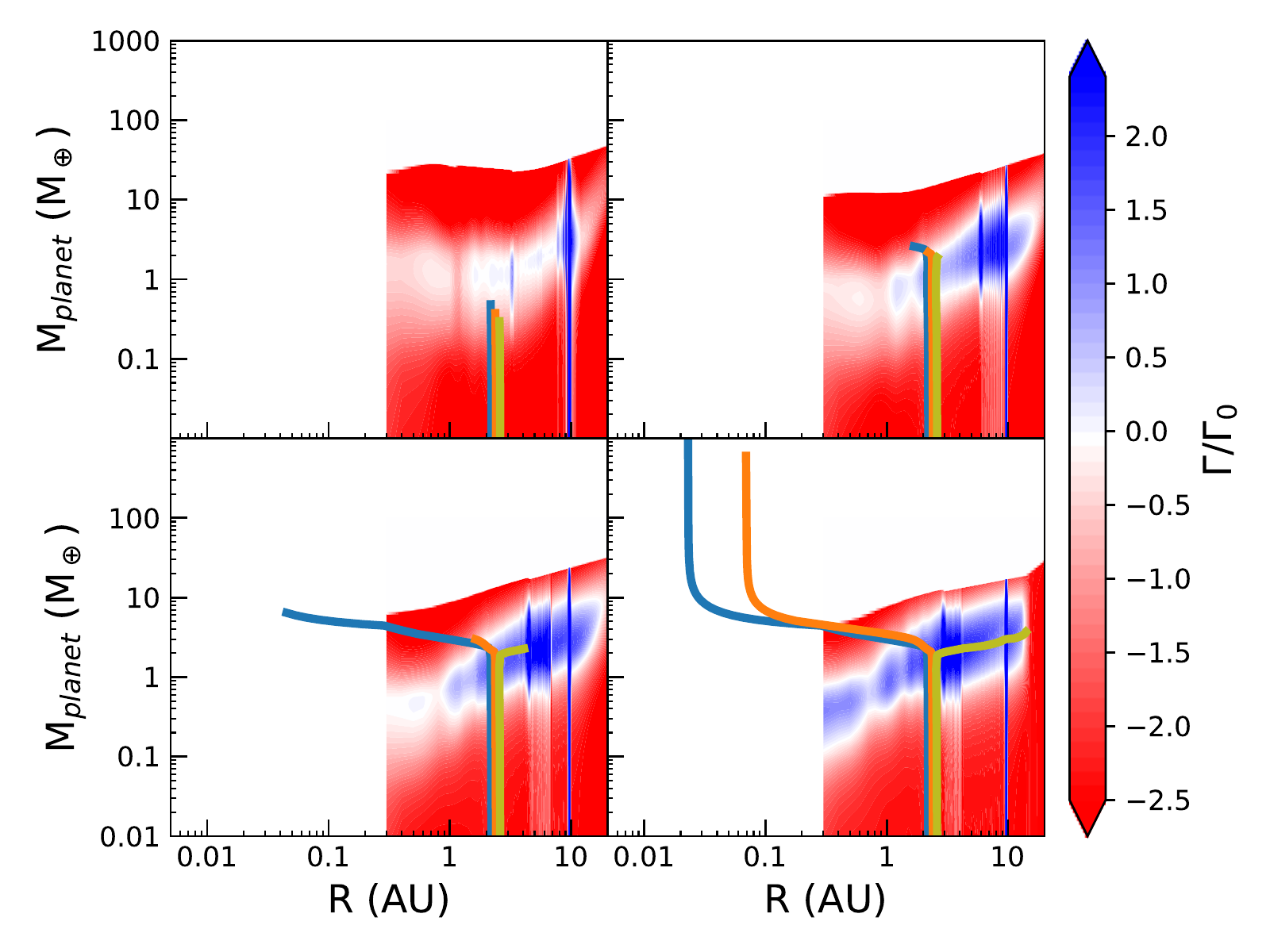}
\put(30,67){\Large a. 0.1 Myr}
\put(63,67){\Large b. 0.5 Myr}
\put(30,34){\Large c. 1.7 Myr}
\put(63,34){\Large d. 4.5 Myr}
\end{overpic}
\caption{ Combined torque maps and planet formation tracks for four different snapshots throughout the formation history of our three planets of interest. In the torque maps (coloured region) red denotes inward migration, white denotes zero net torque, and blue denotes outward migration. The band of outward migration between 1-10 AU is caused by our choice of disk model, and the opacity arising from CO$_2$ freeze out.}
\label{fig:tracks02}
\end{figure*}

In Figure \ref{fig:tracks02} we show the individual formation tracks for the three planets along with a map of the total torque. In the map, red denotes inward migration, white is zero torque, and blue shows outward migration. The large band of outward migration that grows as the disk ages is caused by a combination of our choice of disk model (which steepens with time), and the opacity transition that occurs that the H$_2$O and CO$_2$ ice lines. The mass independent outward migration feature at 10 AU is caused by a discontinuity in the temperature profile caused by our setting a minimum temperature of 20 K for radii outside of 10 AU. The colour map extends up to the gap opening mass, where Type-I torques cease to be relevant for planetary migration.

In Figure \ref{fig:tracks02}a we show an early snapshot of the formation tracks. Up to this point, the planets are not sufficiently massive to show significant migration, hence the tracks grow almost completely in the vertical direction.

Once the planets reach a mass of a few M$_\oplus$ they begin to feel the effect of the disk torques. In Figure \ref{fig:tracks02}b the blue planet has grown so massive that it misses the band of outward migration, and begins to rapidly move inward. The orange planet has coincidentally grown to a mass where it sits in a region of zero torque, and hence has not migrated far in the first 0.5 Myr. The green planet's growth places it in the outward migrating band, and hence it begins to migrate outward.

By Figure \ref{fig:tracks02}c the blue planet has opened a gap (decoupling it from Type-I migrating torques) and its core has nearly reached its final mass. At this stage its gas accretion is limited by eq. \ref{eq:math03} and it is sufficiently massive that Type-II migration is also suppressed. The orange planet has grown enough to escape the region of zero torque that it has previously occupied, and begins to migrate inwards rapidly. Meanwhile the green planet continues to migrate outwards. Since it continues to move to larger radii, and smaller gas / solid surface densities its growth is suppressed when compared to the other planets which move inward, to higher surface densities.

In Figure \ref{fig:tracks02}d we show a late time in the evolution of the disk. At this point the orange planet has opened its gap and accreted nearly all of its available material. The green planet reaches the end of the outward migration band and becomes trapped between inward and outward migration. Once it grows enough to open a gap in the disk it will begin to move inward under Type-II migration.

\subsection{Carbon-to-Oxygen ratio}

\begin{figure*}
\centering
\begin{overpic}[width=\textwidth]{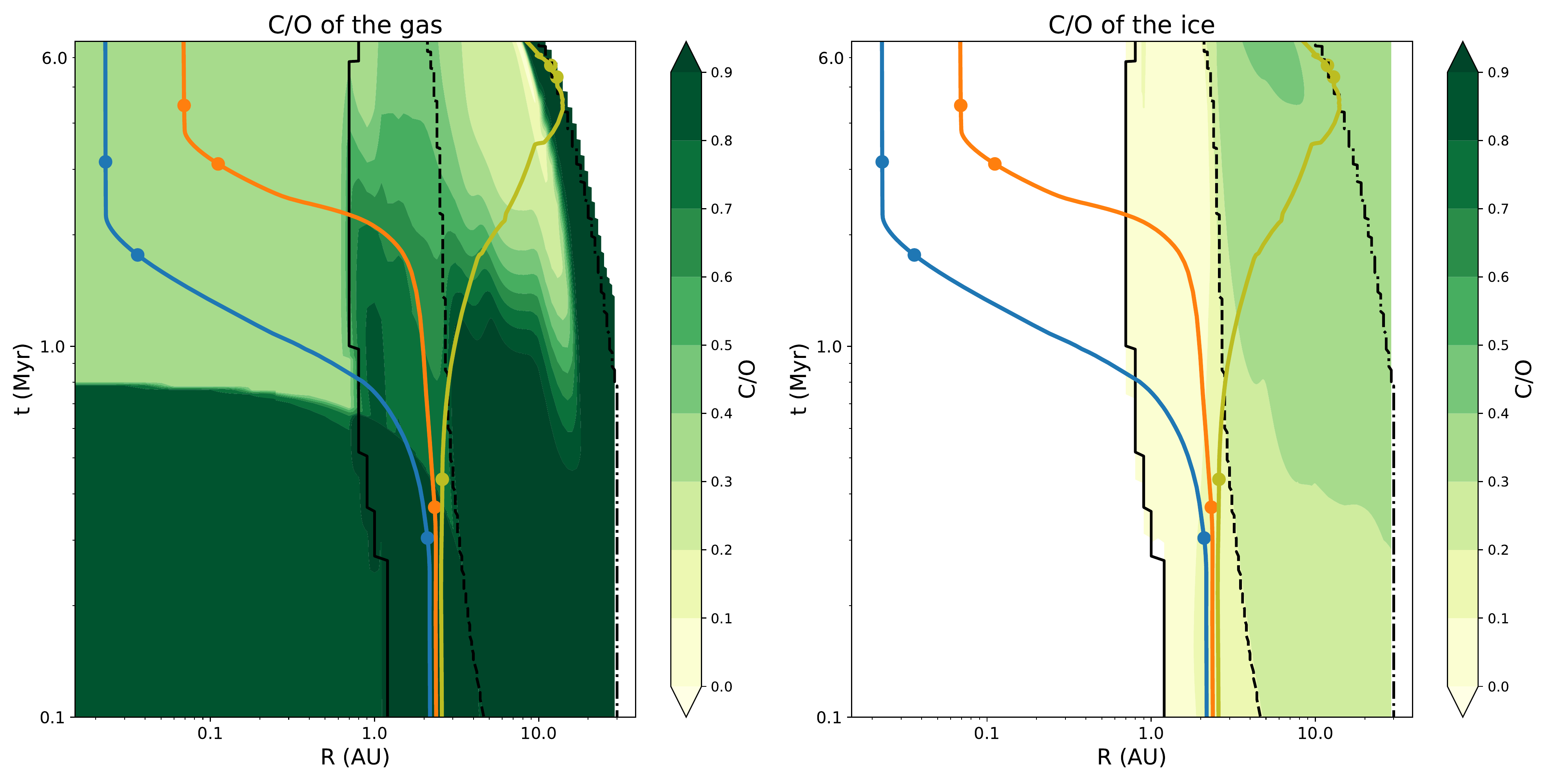}
\put(18,40){H$_2$O}
\put(28,40){CO$_2$}
\put(37,40){CO}
\end{overpic}
\caption{Similar to Figure \ref{fig:chem01}, but here we also include the effect of the excess gaseous carbon caused by the `reset' scenario of carbon depletion. We see that it advects inward quickly, disappearing before 1 Myr. The `ongoing' scenario would maintain the same level of carbon excess over the lifetime of the disk. Additionally we over-plot the radial evolution of the three targeted planets. The dots along each track denote the point where (starting from the bottom); the planet begins accreting gas, when it has accreted 3 M$_\oplus$ of gas, and when the planet has reached half of its final mass. The inner two planets accrete their gas primarily within the ice line, while the outer planet accretes its gas between the CO$_2$ and CO ice lines. As a result it probes the volatile chemical evolution that occurs in that region of the disk.}
\label{fig:tracks03}
\end{figure*}

In Figure \ref{fig:tracks03} we have over-plotted the evolving map of C/O with the orbital radius evolution for the three planets along with excess C/O that is generated by the `reset' scenario of refractory carbon destruction (dark green region within 5 AU). In the case of the `ongoing' scenario, the dark green region of the figure would extend for the whole life of the disk, while in the `reset' scenario it eventually vanishes. Here we illustrate the chemical implication of the migration history of the planets. Along the tracks we have denoted three important milestones with points (starting from the bottom): the point when the planet begins accreting gas, when it has accreted 3 M$_\oplus$ of gas, and when it has accreted half of its mass. 

While all three of the planets begin their gas accretion at nearly the same time, their migration histories can end up delaying when they hit the later milestones. For example: the blue and orange planets both end up in a similar mass regime, but since the orange planet's early migration was slower than the blue planet, it did not reach its half mass until $\sim 1$ Myr later. Similarly, since the green planet underwent significant outward migration, the point where it reached 3 M$_\oplus$ of gas was delayed by up to $\sim 3$ Myr relative to the two other planets.

When and where planets accrete their material has important implications for the chemical make-up of their atmosphere. Recall that we used a simple prescription for determining if refractories are delivered to the planetary atmosphere during a planetesimal collision. When the atmosphere is heavier than 3 M$_\oplus$ the planetesimal is completely destroyed, releasing its entire abundance of refractories and volatiles. Below this cut off, only the volatiles are delivered to the atmosphere.

\subsubsection{Hot Jupiters}

Both blue and orange planets accrete the majority of their material inside the water ice line where the gas is significantly oxygen rich (in the case of the `reset' carbon excess scenario). Additionally they accrete well within the part of the disk where the solids are carbon depleted, hence their refractory accretion will also be oxygen rich. The only significant contribution to the carbon inventory can come from the carbon rich gas which is produced through one of the solid carbon depletion mechanisms discussed before. 

This is illustrated by including our carbon excess models to the evolution map of the gaseous C/O. In Figure \ref{fig:tracks03} we show the effect of the `reset' scenario, where the gaseous carbon excess is produced early in the disk lifetime, then advects towards the host star with the bulk gas. In this case the carbon rich gas quickly (within 1 Myr) vanishes, leaving behind the oxygen rich gas we would have had if carbon enrichment was not considered. In the `ongoing' scenario, the carbon excess does not disappear, maintaining the carbon rich gas for the entire lifetime of the disk.

\begin{figure}
\centering
\begin{overpic}[width=0.5\textwidth]{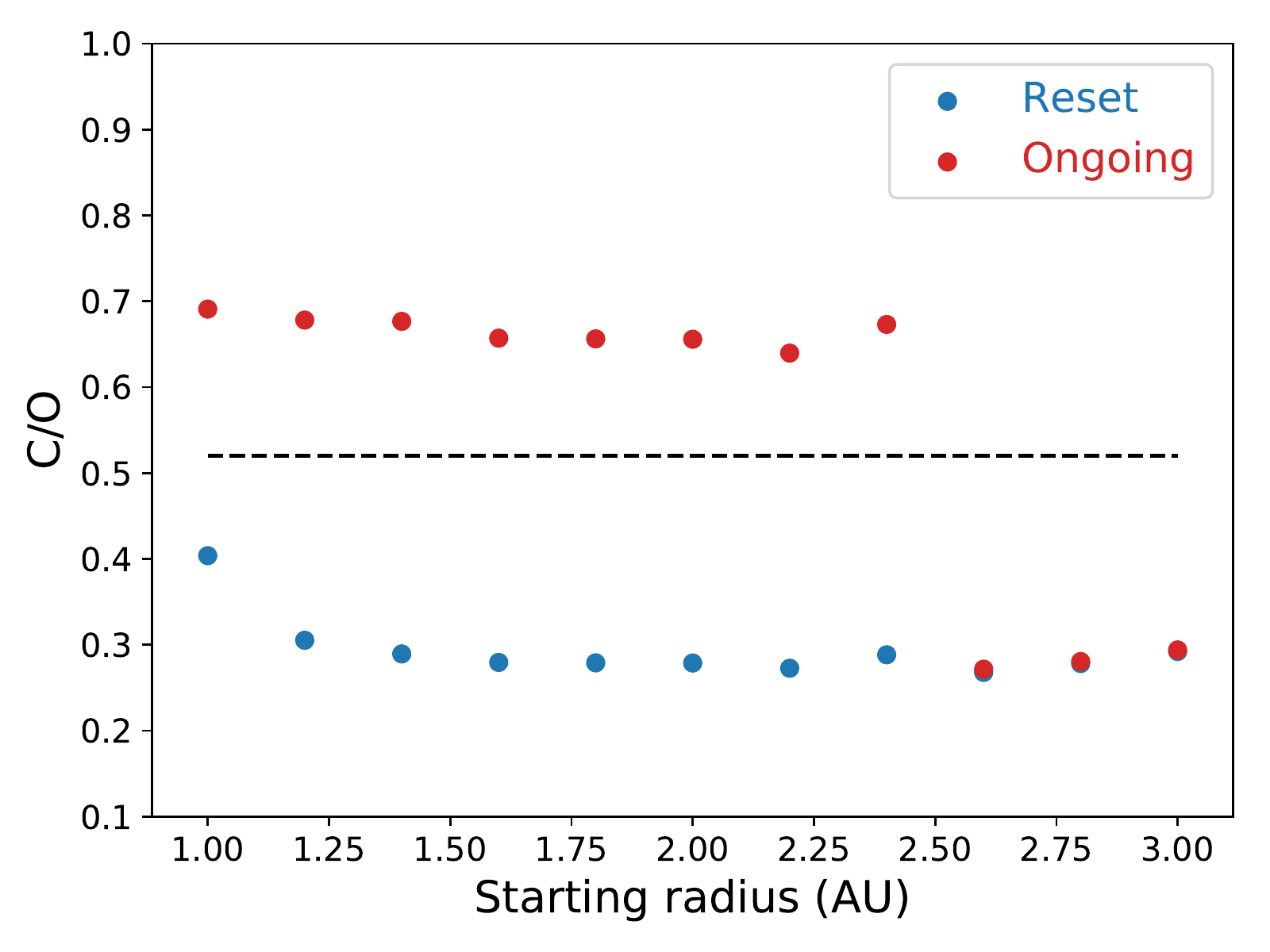}
\put(75,41){ Stellar C/O }
\end{overpic}
\caption{Carbon-to-oxygen ratio (C/O) for the atmospheres of the synthetic planets. In red we plot the case where carbon depletion is ongoing, and the carbon excess in the gas is maintained at the same abundance throughout the whole disk life. In blue we plot the case where carbon depletion occurred in a single `reset' event near the beginning of the disk life. In this case the carbon excess in the gas would advect inward with the accreting gas. Included is the stellar C/O (dashed line).}
\label{fig:CtoO01}
\end{figure}

In Figure \ref{fig:CtoO01} the resulting C/O for the atmosphere of each of the synthetic planets and for each of the carbon excess models is presented. Generally speaking in the `ongoing' excess model the planets end up more carbon-rich than in the `reset' model. This is not surprising because each of the planets accrete the majority of their gas at times $> 1$ Myr at which point the carbon excess from the `reset' model has accreted into the host star. In effect, the results of the `reset' model are indistinguishable from a result of not including the excess gaseous carbon at all, as was done in \cite{Mordasini16}. 

The difference is minimal in the planets with initial orbital radii larger than 2.4 AU. These planets accrete the majority of their gas outside of 5 AU, which we assume is the outer radius of the carbon depleted region (\S \ref{sec:refdepl}). Hence the C/O of these Neptune-sized planets is independent of the particular carbon excess model. 

The hot Jupiters that formed in the `reset' model (starting at radii $< 2.4$ AU) all ended with C/O that would be lower than their host star ($\sim 0.52$). This is driven by the fact that the carbon excess component of the gas disappears within 1 Myr (recall Figure \ref{fig:tracks03}), prior to when the hot Jupiters accrete the majority of their gas. In the `ongoing' scenario the gaseous carbon excess contributes strongly to the bulk carbon abundance in the planetary atmosphere, resulting in carbon masses that increase by a factor of a few times the carbon accreted from the gas in the `reset' scenario (see Table \ref{tab:res02}). As a result these planets end their accretion phase with an atmosphere that is around twice as carbon rich relative to oxygen as the planets formed in the `reset' scenario. 

\begin{table*}
\centering
\caption{Total carbon and oxygen mass inventory from volatile gas (g), ice (i), and refractory (gr) accretion in the planetary atmosphere. Note that this does not include refractories that survived to the core, nor any mixing between the core and the atmosphere. For the gaseous carbon component we list the resulting abundances from both the reset (r) and ongoing (o) models. Where $X(Y)$ is $X\times 10^Y$. We additionally list the elemental ratios for each of the planetary atmosphere plotted in Figure \ref{fig:CtoO01}.}
\ignore{
\begin{tabular}{| c | c | c | c | c |}
Initial radius & C(g) & O(g) & C(s) & O(s) \\\hline
(AU) & (M$_\oplus$) - r/o & (M$_\oplus$) & (M$_\oplus$) & (M$_\oplus$) & (M$_\oplus$) & (M$_\oplus$)  \\\hline
 1.0 & 1.28 / 2.78 & 1.40 & 0.40 & 3.46 \\\hline
 1.2 & 0.82 / 2.80 & 1.45 & 0.46 & 3.90 \\\hline
 1.4 & 0.75 / 2.80 & 1.46 & 0.48 & 3.98 \\\hline
 1.6 & 0.72 / 2.78 & 1.42 & 0.59 & 4.66 \\\hline
 1.8 & 0.70 / 2.74 & 1.37 & 0.60 & 4.67 \\\hline
 2.0 & 0.66 / 2.70 & 1.30 & 0.63 & 4.72 \\\hline
 2.2 & 0.59 / 2.66 & 1.17 & 0.77 & 5.41 \\\hline
 2.4 & 0.18 / 1.67 & 0.37 & 0.51 & 2.84 \\\hline
 2.6 & 0.0074 / 0.0076 & 0.039 & 0.0039 & 0.0018 \\\hline
 2.8 & 0.0023 / 0.0025 & 0.012 & 0.0020 & 0.0002 \\\hline
 3.0 & 0.0009 / 0.0009 & 0.004 & 0.0011 & 4$\times 10^{-7}$ \\\hline
\end{tabular}
}
\setlength{\tabcolsep}{4.5pt}
\begin{tabular}{| c | c | c | c | c | c | c | c | c | c | c |}
\hline
r$_0$ & C(g) & O(g) & C(i) & O(i) & C(gr) & O(gr) & \multicolumn{4}{c|}{Elemental Ratios} \\\hline
AU & M$_\oplus$ - r/o & M$_\oplus$ & M$_\oplus$ & M$_\oplus$ & M$_\oplus$ & M$_\oplus$ & C/O$_{\rm r}$ & C/O$_{\rm o}$ & C/N$_{\rm r}$ & C/N$_{\rm o}$ \\\hline
 1.0 & 2.11(0) / 3.61(0) & 5.88(0) & 2.8(-9) & 4.9(-7) & 1.9(-4) & 1.08(0) & 0.40 & 0.69 & 4.0 & 6.8 \\\hline
 1.2 & 1.62(0) / 3.59(0) & 5.85(0) & 3.3(-8) & 1.3(-5) & 2.1(-4) & 1.21(0) & 0.31 & 0.68 & 3.1 & 6.8 \\\hline
 1.4 & 1.54(0) / 3.60(0) & 5.86(0) & 1.2(-5) & 4.3(-3) & 2.1(-4) & 1.23(0) & 0.29 & 0.68 & 2.9 & 6.8 \\\hline
 1.6 & 1.53(0) / 3.59(0) & 5.85(0) & 6.9(-5) & 2.5(-2) & 2.5(-4) & 1.42(0) & 0.28 & 0.66 & 2.9 & 6.8 \\\hline
 1.8 & 1.51(0) / 3.55(0) & 5.78(0) & 9.8(-5) & 3.1(-2) & 2.4(-4) & 1.40(0) & 0.28 & 0.66 & 2.9 & 6.8 \\\hline
 2.0 & 1.51(0) / 3.55(0) & 5.78(0) & 4.7(-4) & 5.5(-2) & 2.4(-4) & 1.38(0) & 0.28 & 0.65 & 2.9 & 6.8 \\\hline
 2.2 & 1.53(0) / 3.60(0) & 5.86(0) & 4.9(-3) & 9.0(-2) & 2.7(-4) & 1.55(0) & 0.27 & 0.64 & 2.9 & 6.8 \\\hline
 2.4 & 1.10(0) / 2.58(0) & 4.21(0) & 9.8(-3) & 1.7(-1) & 1.3(-4) & 7.5(-1) & 0.29 & 0.67 & 2.9 & 6.8 \\\hline
 2.6 & 3.8(-4) / 6.4(-4) & 1.1(-3) & 1.7(-2) & 9.1(-2) & 1.9(-3) & 1.8(-3) & 0.27 & 0.27 & 4.4 & 4.5 \\\hline
 2.8 & 1.3(-4) / 2.6(-4) & 3.2(-4) & 1.4(-2) & 7.0(-2) & 2.2(-4) & 2.1(-4) & 0.28 & 0.28 & 6.2 & 6.2 \\\hline
 3.0 & 9.7(-5) / 1.7(-4) & 5.3(-4) & 1.2(-2) & 5.4(-2) & 0.0     & 0.0     & 0.29 & 0.29 & 7.7 & 7.7 \\\hline
\end{tabular}
\label{tab:res02}
\end{table*}

These carbon-rich planets depend strongly on both the treatment of the solid accretion as well as the distribution of the carbon that is processed off of the dust, hence these results should be interpreted with caution. As previously mentioned, the process which drives carbon depletion is not well understood. In the model of \cite{Anderson2017} (which we take as motivation for the `ongoing' model) the dust processing is done above the midplane, in a region of the disk that is less shielded from high energy radiation. The assumption is that the excess carbon is mixed back down to near the midplane where it can be accreted by the forming planet. This dust processing has recently been called into question by \cite{Klarmann2018} who separately compute the radial and vertical transport of the dust grains that are responsible for this carbon depletion. They find that refractory carbon depletion along the midplane is suppressed when considering the combined effect of the grain transport and chemistry.

Given that the current view is that late stage (post-gap opening) gas accretion is done vertically, from above the midplane means that the excess carbon does not need to mix completely to the midplane to be incorporated into the planet. However it is not clear whether the height at which the carbon is produced coincides well with the feeding height of the accreting planet.

In Table \ref{tab:res02} we break down the total carbon and oxygen mass into the contributions made by gas accretion and solid accretion (including ice and refractories). Because the refractories are oxygen rich within 5 AU, the solid contribution to atmospheric oxygen outweighs carbon by almost an order of magnitude in the Hot Jupiter planets. These planets accreted the majority of their atmosphere inside the water ice line (recall figure \ref{fig:tracks03}), so we can be sure that the accretion of ices have had minimal impact on C/O of these planets. We can see that the carbon mass directly accreted from the disk gas into the Hot Jupiter varies by a factor of about 2 depending on our choice of carbon excess model.

\subsubsection{Super-Earths}

The atmospheric C/O for the low mass planets are primarily dominated by the accretion of icy planetesimals. They tend to have lower-mass atmospheres for longer, and hence incoming planetesimals tend to deposit their refractory mass directly onto the core of the planet rather than evaporate in the atmosphere. As a result, their low C/O is caused by their accretion of carbon-depleted gas and oxygen rich ice after 1 Myr, driven by the surface chemistry reactions discussed earlier. Note that because of their lower masses, their global C/O may be susceptible to chemical processing of the planetary core, which we ignore.

\section{ Conclusions }\label{sec:con}

How the astrochemisty of protoplanetary disks are imprinted into the atmospheric chemistry of exoplanets is a complex, time-dependent problem that requires a combination of detailed planet formation and disk chemistry models. Here we have combined two such models along with a new treatment of refractory carbon depletion, and its corresponding gaseous carbon excess, to compute the carbon-to-oxygen ratio (C/O) in the gas that forms the atmospheres of synthetic exoplanets.

A particularly important feature of our model is that solar system solids show signs of refractory carbon depletion relative to the carbon observed in the Interstellar medium (ISM). We argue that this missing carbon would have been released into the gas, where it is possibly available for accretion into the atmospheres of growing planets. What dictates the availability of excess carbon depends on the mechanism which produces the carbon depletion, which is currently unknown.

We posit two possible models: `reset' and `ongoing', the first one representing a catastrophic thermal event that occurred very early in the disk lifetime, while the second represents an ongoing process that is constantly producing an excess of gaseous carbon. The excess carbon from the `reset' model advects into the host star within 0.8 Myr, and hence is not accreted onto forming planets, while the `ongoing' model maintains a carbon excess in the inner ($<$ 5 AU) solar system, allowing for carbon-rich (relative to solar) gas to be accreted by forming planets.

This carbon excess model is combined with the results from the chemcial evolution models by \cite{Eistrup2017} to compute the C/O distribution of the gas and solids (ice $+$ dust) over the whole lifetime of the disk. In this evolving astrochemical disk we compute the growth and migration of planets beginning in a range of radii between 1-3 AU. More than half the planets end up as Hot Jupiters, while the others remain super-Earth sized at radii $\geq$ 7 AU, depending on their migration histories. 

In the Hot Jupiters the atmospheric C/O depends on our treatment of the carbon excess model, because their gas accretion is exclusively done in the region where carbon depletion is found in our solar system (volatile ices do not contribute). On the other hand, the super-Earths are more dependent on the details of the chemical model, because the majority of their growth occurs outside the carbon depletion zone of the disk. For the Hot Jupiters we can produce atmospheric C/O that exceed their host star, as long as they accrete gas in the `ongoing' model of carbon excess. Otherwise their C/O is strictly lower than their host star because they accrete gas and solids that are both oxygen rich with respect to the host star.

We have intentionally kept the source of the carbon excess simple, studying only two possible extremes. We have assumed that similar carbon depletion processes could occur in other solar systems, and have not explored variance in initial C/O in both the host star and gas disk. Instead opting to explore what variation in C/O could occur simply from processes internal to the protoplanetary disk. Regardless we have demonstrated that if the general trend of \cite{Brewer16} hold, that most Hot Jupiters have super-stellar C/O, then it is suggestive that there is a source of carbon excess in the inner parts of most stellar system that persists throughout the lifetime of the disk, or at least long enough to be incorporated into the planetary atmospheres.

\begin{acknowledgements}

 Astrochemistry in Leiden is supported by the European Union A-ERC grant 291141
CHEMPLAN, by the Netherlands Research School for Astronomy (NOVA), and
by a Royal Netherlands Academy of Arts and Sciences (KNAW) professor prize.

\end{acknowledgements}

\bibliographystyle{aa} 
\bibliography{mybib.bib} 

\begin{thebibliography}{74}
\expandafter\ifx\csname natexlab\endcsname\relax\def\natexlab#1{#1}\fi

\bibitem[{{Alessi} \& {Pudritz}(2018)}]{Alessi16b}
{Alessi}, M. \& {Pudritz}, R.~E. 2018, ArXiv e-prints
  [\eprint[arXiv]{1804.01148}]

\bibitem[{{Alessi} {et~al.}(2017){Alessi}, {Pudritz}, \& {Cridland}}]{Alessi16}
{Alessi}, M., {Pudritz}, R.~E., \& {Cridland}, A.~J. 2017, \mnras, 464, 428

\bibitem[{{Ali-Dib}(2017)}]{AliDib2017}
{Ali-Dib}, M. 2017, \mnras, 464, 4282

\bibitem[{{Alibert} {et~al.}(2013){Alibert}, {Carron}, {Fortier}, {Pfyffer},
  {Benz}, {Mordasini}, \& {Swoboda}}]{Alibert2013}
{Alibert}, Y., {Carron}, F., {Fortier}, A., {et~al.} 2013, \aap, 558, A109

\bibitem[{{All{\`e}gre} {et~al.}(2001){All{\`e}gre}, {Manh{\`e}s}, \&
  {Lewin}}]{Allegre2001}
{All{\`e}gre}, C., {Manh{\`e}s}, G., \& {Lewin}, {\'E}. 2001, Earth and
  Planetary Science Letters, 185, 49

\bibitem[{{Anderson} {et~al.}(2017){Anderson}, {Bergin}, {Blake}, {Ciesla},
  {Visser}, \& {Lee}}]{Anderson2017}
{Anderson}, D.~E., {Bergin}, E.~A., {Blake}, G.~A., {et~al.} 2017, \apj, 845,
  13

\bibitem[{{Atreya} {et~al.}(2016){Atreya}, {Crida}, {Guillot}, {Lunine},
  {Madhusudhan}, \& {Mousis}}]{Atreya2016}
{Atreya}, S.~K., {Crida}, A., {Guillot}, T., {et~al.} 2016, ArXiv e-prints
  [\eprint[arXiv]{1606.04510}]

\bibitem[{{Batygin}(2018)}]{Batygin2018}
{Batygin}, K. 2018, \aj, 155, 178

\bibitem[{{Bergin} {et~al.}(2015){Bergin}, {Blake}, {Ciesla}, {Hirschmann}, \&
  {Li}}]{Berg15}
{Bergin}, E.~A., {Blake}, G.~A., {Ciesla}, F., {Hirschmann}, M.~M., \& {Li}, J.
  2015, Proceedings of the National Academy of Science, 112, 8965

\bibitem[{{Bitsch} {et~al.}(2015){Bitsch}, {Lambrechts}, \&
  {Johansen}}]{Bitsch2015}
{Bitsch}, B., {Lambrechts}, M., \& {Johansen}, A. 2015, \aap, 582, A112

\bibitem[{{Bitsch} {et~al.}(2018){Bitsch}, {Morbidelli}, {Johansen}, {Lega},
  {Lambrechts}, \& {Crida}}]{Bitsch2018}
{Bitsch}, B., {Morbidelli}, A., {Johansen}, A., {et~al.} 2018, \aap, 612, A30

\bibitem[{{Boogert} {et~al.}(2015){Boogert}, {Gerakines}, \&
  {Whittet}}]{Boogert2015}
{Boogert}, A.~C.~A., {Gerakines}, P.~A., \& {Whittet}, D.~C.~B. 2015, \araa,
  53, 541

\bibitem[{{Booth} {et~al.}(2017){Booth}, {Clarke}, {Madhusudhan}, \&
  {Ilee}}]{Booth2017}
{Booth}, R.~A., {Clarke}, C.~J., {Madhusudhan}, N., \& {Ilee}, J.~D. 2017,
  \mnras, 469, 3994

\bibitem[{{Bosman} {et~al.}(2017){Bosman}, {Tielens}, \& {van
  Dishoeck}}]{Bosman2017b}
{Bosman}, A.~D., {Tielens}, A.~G.~G.~M., \& {van Dishoeck}, E.~F. 2017, ArXiv
  e-prints [\eprint[arXiv]{1712.03989}]

\bibitem[{{Brewer} {et~al.}(2017){Brewer}, {Fischer}, \&
  {Madhusudhan}}]{Brewer16}
{Brewer}, J.~M., {Fischer}, D.~A., \& {Madhusudhan}, N. 2017, \aj, 153, 83

\bibitem[{{Brouwers} {et~al.}(2018){Brouwers}, {Vazan}, \&
  {Ormel}}]{Brouwers2018}
{Brouwers}, M.~G., {Vazan}, A., \& {Ormel}, C.~W. 2018, \aap, 611, A65

\bibitem[{{Christensen} {et~al.}(2009){Christensen}, {Holzwarth}, \&
  {Reiners}}]{Christensen2009}
{Christensen}, U.~R., {Holzwarth}, V., \& {Reiners}, A. 2009, \nat, 457, 167

\bibitem[{{Cleeves} {et~al.}(2014){Cleeves}, {Bergin}, \& {Adams}}]{Cleeves14}
{Cleeves}, L.~I., {Bergin}, E.~A., \& {Adams}, F.~C. 2014, \apj, 794, 123

\bibitem[{{Coleman} \& {Nelson}(2014)}]{Cole14}
{Coleman}, G.~A.~L. \& {Nelson}, R.~P. 2014, \mnras, 445, 479

\bibitem[{{Coleman} \& {Nelson}(2016)}]{Cole16}
{Coleman}, G.~A.~L. \& {Nelson}, R.~P. 2016, \mnras, 460, 2779

\bibitem[{{Cridland}(2018)}]{Crid2018}
{Cridland}, A.~J. 2018, ArXiv e-prints [\eprint[arXiv]{1809.04657}]

\bibitem[{{Cridland} {et~al.}(2016){Cridland}, {Pudritz}, \&
  {Alessi}}]{Crid16a}
{Cridland}, A.~J., {Pudritz}, R.~E., \& {Alessi}, M. 2016, \mnras, 461, 3274

\bibitem[{{Cridland} {et~al.}(2017{\natexlab{a}}){Cridland}, {Pudritz}, \&
  {Birnstiel}}]{Crid16b}
{Cridland}, A.~J., {Pudritz}, R.~E., \& {Birnstiel}, T. 2017{\natexlab{a}},
  \mnras, 465, 3865

\bibitem[{{Cridland} {et~al.}(2017{\natexlab{b}}){Cridland}, {Pudritz},
  {Birnstiel}, {Cleeves}, \& {Bergin}}]{Crid17}
{Cridland}, A.~J., {Pudritz}, R.~E., {Birnstiel}, T., {Cleeves}, L.~I., \&
  {Bergin}, E.~A. 2017{\natexlab{b}}, \mnras, 469, 3910

\bibitem[{{Dittkrist} {et~al.}(2014){Dittkrist}, {Mordasini}, {Klahr},
  {Alibert}, \& {Henning}}]{D14}
{Dittkrist}, K.-M., {Mordasini}, C., {Klahr}, H., {Alibert}, Y., \& {Henning},
  T. 2014, \aap, 567, A121

\bibitem[{{Drozdovskaya} {et~al.}(2016){Drozdovskaya}, {Walsh}, {van Dishoeck},
  {Furuya}, {Marboeuf}, {Thiabaud}, {Harsono}, \& {Visser}}]{Drozd16}
{Drozdovskaya}, M.~N., {Walsh}, C., {van Dishoeck}, E.~F., {et~al.} 2016,
  \mnras, 462, 977

\bibitem[{{Drozdovskaya} {et~al.}(2014){Drozdovskaya}, {Walsh}, {Visser},
  {Harsono}, \& {van Dishoeck}}]{Drozd14}
{Drozdovskaya}, M.~N., {Walsh}, C., {Visser}, R., {Harsono}, D., \& {van
  Dishoeck}, E.~F. 2014, \mnras, 445, 913

\bibitem[{{Eistrup} {et~al.}(2016){Eistrup}, {Walsh}, \& {van
  Dishoeck}}]{Eistrup2016}
{Eistrup}, C., {Walsh}, C., \& {van Dishoeck}, E.~F. 2016, \aap, 595, A83

\bibitem[{{Eistrup} {et~al.}(2018){Eistrup}, {Walsh}, \& {van
  Dishoeck}}]{Eistrup2017}
{Eistrup}, C., {Walsh}, C., \& {van Dishoeck}, E.~F. 2018, \aap, 613, A14

\bibitem[{{Fogel} {et~al.}(2011){Fogel}, {Bethell}, {Bergin}, {Calvet}, \&
  {Semenov}}]{Fogel11}
{Fogel}, J.~K.~J., {Bethell}, T.~J., {Bergin}, E.~A., {Calvet}, N., \&
  {Semenov}, D. 2011, \apj, 726, 29

\bibitem[{{Hasegawa} \& {Pudritz}(2011)}]{HP11}
{Hasegawa}, Y. \& {Pudritz}, R.~E. 2011, \mnras, 417, 1236

\bibitem[{{Hasegawa} \& {Pudritz}(2013)}]{HP13}
{Hasegawa}, Y. \& {Pudritz}, R.~E. 2013, \apj, 778, 78

\bibitem[{{Hasegawa} \& {Pudritz}(2014)}]{HP14}
{Hasegawa}, Y. \& {Pudritz}, R.~E. 2014, \apj, 794, 25

\bibitem[{{Horn} {et~al.}(2012){Horn}, {Lyra}, {Mac Low}, \&
  {S{\'a}ndor}}]{Horn2012}
{Horn}, B., {Lyra}, W., {Mac Low}, M.-M., \& {S{\'a}ndor}, Z. 2012, \apj, 750,
  34

\bibitem[{{Ida} \& {Lin}(2004)}]{IL04a}
{Ida}, S. \& {Lin}, D.~N.~C. 2004, \apj, 604, 388

\bibitem[{{Ida} \& {Lin}(2008)}]{IdaLin2008}
{Ida}, S. \& {Lin}, D.~N.~C. 2008, \apj, 673, 487

\bibitem[{{Ikoma} {et~al.}(2000){Ikoma}, {Nakazawa}, \& {Emori}}]{Ikoma2000}
{Ikoma}, M., {Nakazawa}, K., \& {Emori}, H. 2000, \apj, 537, 1013

\bibitem[{{Jura}(2008)}]{Jura2008}
{Jura}, M. 2008, \aj, 135, 1785

\bibitem[{{Klarmann} {et~al.}(2018){Klarmann}, {Ormel}, \&
  {Dominik}}]{Klarmann2018}
{Klarmann}, L., {Ormel}, C.~W., \& {Dominik}, C. 2018, ArXiv e-prints
  [\eprint[arXiv]{1809.01648}]

\bibitem[{{Kokubo} \& {Ida}(2002)}]{KI02}
{Kokubo}, E. \& {Ida}, S. 2002, \apj, 581, 666

\bibitem[{{Lambrechts} \& {Johansen}(2014)}]{LambJoh2014}
{Lambrechts}, M. \& {Johansen}, A. 2014, \aap, 572, A107

\bibitem[{{Lee} {et~al.}(2010){Lee}, {Bergin}, \& {Nomura}}]{Lee2010}
{Lee}, J.-E., {Bergin}, E.~A., \& {Nomura}, H. 2010, \apjl, 710, L21

\bibitem[{{Lyra} {et~al.}(2010){Lyra}, {Paardekooper}, \& {Mac Low}}]{Lyra2010}
{Lyra}, W., {Paardekooper}, S.-J., \& {Mac Low}, M.-M. 2010, \apjl, 715, L68

\bibitem[{{Madhusudhan} {et~al.}(2014){Madhusudhan}, {Amin}, \&
  {Kennedy}}]{Madu2014}
{Madhusudhan}, N., {Amin}, M.~A., \& {Kennedy}, G.~M. 2014, \apjl, 794, L12

\bibitem[{{Madhusudhan} {et~al.}(2017){Madhusudhan}, {Bitsch}, {Johansen}, \&
  {Eriksson}}]{Madhu2017}
{Madhusudhan}, N., {Bitsch}, B., {Johansen}, A., \& {Eriksson}, L. 2017,
  \mnras, 469, 4102

\bibitem[{{Masset} {et~al.}(2006){Masset}, {Morbidelli}, {Crida}, \&
  {Ferreira}}]{Masset06}
{Masset}, F.~S., {Morbidelli}, A., {Crida}, A., \& {Ferreira}, J. 2006, \apj,
  642, 478

\bibitem[{{Matsumura} \& {Pudritz}(2006)}]{MP06}
{Matsumura}, S. \& {Pudritz}, R.~E. 2006, \mnras, 365, 572

\bibitem[{{McElroy} {et~al.}(2013){McElroy}, {Walsh}, {Markwick}, {Cordiner},
  {Smith}, \& {Millar}}]{McE03}
{McElroy}, D., {Walsh}, C., {Markwick}, A.~J., {et~al.} 2013, \aap, 550, A36

\bibitem[{{Mishra} \& {Li}(2015)}]{MishraLi2015}
{Mishra}, A. \& {Li}, A. 2015, \apj, 809, 120

\bibitem[{{Miyake} \& {Nakagawa}(1993)}]{Miyake1993}
{Miyake}, K. \& {Nakagawa}, Y. 1993, \icarus, 106, 20

\bibitem[{{Morbidelli} {et~al.}(2014){Morbidelli}, {Szul{\'a}gyi}, {Crida},
  {Lega}, {Bitsch}, {Tanigawa}, \& {Kanagawa}}]{Morbidelli2014}
{Morbidelli}, A., {Szul{\'a}gyi}, J., {Crida}, A., {et~al.} 2014, \icarus, 232,
  266

\bibitem[{{Mordasini}(2013)}]{Mordasini13}
{Mordasini}, C. 2013, \aap, 558, A113

\bibitem[{{Mordasini}(2014)}]{Mordasini14b}
{Mordasini}, C. 2014, \aap, 572, A118

\bibitem[{{Mordasini} {et~al.}(2006){Mordasini}, {Alibert}, \&
  {Benz}}]{Mordasini2006}
{Mordasini}, C., {Alibert}, Y., \& {Benz}, W. 2006, in Tenth Anniversary of 51
  Peg-b: Status of and prospects for hot Jupiter studies, ed. L.~{Arnold},
  F.~{Bouchy}, \& C.~{Moutou}, 84--86

\bibitem[{{Mordasini} {et~al.}(2011){Mordasini}, {Dittkrist}, {Alibert},
  {Klahr}, {Benz}, \& {Henning}}]{Mordasini2011}
{Mordasini}, C., {Dittkrist}, K.-M., {Alibert}, Y., {et~al.} 2011, in IAU
  Symposium, Vol. 276, The Astrophysics of Planetary Systems: Formation,
  Structure, and Dynamical Evolution, ed. A.~{Sozzetti}, M.~G. {Lattanzi}, \&
  A.~P. {Boss}, 72--75

\bibitem[{{Mordasini} {et~al.}(2015){Mordasini}, {Molli{\`e}re}, {Dittkrist},
  {Jin}, \& {Alibert}}]{Mordasini15}
{Mordasini}, C., {Molli{\`e}re}, P., {Dittkrist}, K.-M., {Jin}, S., \&
  {Alibert}, Y. 2015, International Journal of Astrobiology, 14, 201

\bibitem[{{Mordasini} {et~al.}(2016){Mordasini}, {van Boekel}, {Molli{\`e}re},
  {Henning}, \& {Benneke}}]{Mordasini16}
{Mordasini}, C., {van Boekel}, R., {Molli{\`e}re}, P., {Henning}, T., \&
  {Benneke}, B. 2016, \apj, 832, 41

\bibitem[{{{\"O}berg} {et~al.}(2011){{\"O}berg}, {Murray-Clay}, \&
  {Bergin}}]{Oberg11}
{{\"O}berg}, K.~I., {Murray-Clay}, R., \& {Bergin}, E.~A. 2011, \apjl, 743, L16

\bibitem[{{Ormel} \& {Klahr}(2010)}]{Ormel2010}
{Ormel}, C.~W. \& {Klahr}, H.~H. 2010, \aap, 520, A43

\bibitem[{{Paardekooper} {et~al.}(2011){Paardekooper}, {Baruteau}, \&
  {Kley}}]{Paard11}
{Paardekooper}, S.-J., {Baruteau}, C., \& {Kley}, W. 2011, \mnras, 410, 293

\bibitem[{{Pinilla} {et~al.}(2016){Pinilla}, {Klarmann}, {Birnstiel},
  {Benisty}, {Dominik}, \& {Dullemond}}]{Pinilla2016}
{Pinilla}, P., {Klarmann}, L., {Birnstiel}, T., {et~al.} 2016, \aap, 585, A35

\bibitem[{{Pollack} {et~al.}(1996){Pollack}, {Hubickyj}, {Bodenheimer},
  {Lissauer}, {Podolak}, \& {Greenzweig}}]{Pollack1996}
{Pollack}, J.~B., {Hubickyj}, O., {Bodenheimer}, P., {et~al.} 1996, \icarus,
  124, 62

\bibitem[{{Pontoppidan} {et~al.}(2014){Pontoppidan}, {Salyk}, {Bergin},
  {Brittain}, {Marty}, {Mousis}, \& {{\"O}berg}}]{Pon14}
{Pontoppidan}, K.~M., {Salyk}, C., {Bergin}, E.~A., {et~al.} 2014, Protostars
  and Planets VI, 363

\bibitem[{{Raettig} {et~al.}(2015){Raettig}, {Klahr}, \& {Lyra}}]{Rae15}
{Raettig}, N., {Klahr}, H., \& {Lyra}, W. 2015, \apj, 804, 35

\bibitem[{{Sch{\"a}fer} {et~al.}(2017){Sch{\"a}fer}, {Yang}, \&
  {Johansen}}]{Schafer2017}
{Sch{\"a}fer}, U., {Yang}, C.-C., \& {Johansen}, A. 2017, \aap, 597, A69

\bibitem[{{Schlichting} {et~al.}(2015){Schlichting}, {Sari}, \&
  {Yalinewich}}]{Schlichting2015}
{Schlichting}, H.~E., {Sari}, R., \& {Yalinewich}, A. 2015, \icarus, 247, 81

\bibitem[{{Shakura} \& {Sunyaev}(1973)}]{SS73}
{Shakura}, N.~I. \& {Sunyaev}, R.~A. 1973, \aap, 24, 337

\bibitem[{{Szul{\'a}gyi} {et~al.}(2014){Szul{\'a}gyi}, {Morbidelli}, {Crida},
  \& {Masset}}]{Szul2014}
{Szul{\'a}gyi}, J., {Morbidelli}, A., {Crida}, A., \& {Masset}, F. 2014, \apj,
  782, 65

\bibitem[{{Thiabaud} {et~al.}(2015){Thiabaud}, {Marboeuf}, {Alibert}, {Leya},
  \& {Mezger}}]{Thiabaud2015}
{Thiabaud}, A., {Marboeuf}, U., {Alibert}, Y., {Leya}, I., \& {Mezger}, K.
  2015, \aap, 574, A138

\bibitem[{{Visser} {et~al.}(2011){Visser}, {Doty}, \& {van
  Dishoeck}}]{Visser2011}
{Visser}, R., {Doty}, S.~D., \& {van Dishoeck}, E.~F. 2011, \aap, 534, A132

\bibitem[{{Visser} {et~al.}(2009){Visser}, {van Dishoeck}, {Doty}, \&
  {Dullemond}}]{Visser2009}
{Visser}, R., {van Dishoeck}, E.~F., {Doty}, S.~D., \& {Dullemond}, C.~P. 2009,
  \aap, 495, 881

\bibitem[{{Walsh} {et~al.}(2015){Walsh}, {Nomura}, \& {van Dishoeck}}]{Walsh15}
{Walsh}, C., {Nomura}, H., \& {van Dishoeck}, E. 2015, \aap, 582, A88

\bibitem[{{Weidenschilling}(1977)}]{W77}
{Weidenschilling}, S.~J. 1977, \mnras, 180, 57

\bibitem[{{Whittet} {et~al.}(2013){Whittet}, {Poteet}, {Chiar}, {Pagani},
  {Bajaj}, {Horne}, {Shenoy}, \& {Adamson}}]{Whittet2013}
{Whittet}, D.~C.~B., {Poteet}, C.~A., {Chiar}, J.~E., {et~al.} 2013, \apj, 774,
  102

\end{thebibliography}

\end{document}